\def\h{${{h}_{75}}^{-1}$}

\def\lesssim{\mathrel{\hbox{\rlap{\hbox{\lower4pt\hbox{$\sim$}}}\hbox{$<$}}}}
\def\gtrsim{\mathrel{\hbox{\rlap{\hbox{\lower4pt\hbox{$\sim$}}}\hbox{$>$}}}}
\def\la{\mathrel{\hbox{\rlap{\hbox{\lower4pt\hbox{$\sim$}}}\hbox{$<$}}}}
\def\ga{\mathrel{\hbox{\rlap{\hbox{\lower4pt\hbox{$\sim$}}}\hbox{$>$}}}}

\documentclass{aa}
\usepackage{graphicx}
\usepackage{txfonts} 
\begin{document}

\title {ATCA observations of the galaxy cluster Abell~3921}
\subtitle{I. Radio emission from the central merging sub-clusters}
\author{Ferrari, C.\inst{1} 
\and Hunstead, R.W.\inst{2} 
\and Feretti, L.\inst{3}
\and Maurogordato, S.\inst{4}
\and Schindler, S.\inst{1}}

\offprints{Chiara Ferrari}  
\institute{Institut f\"ur Astro- und Teilchen Physik, Universit\"at Innsbruck,
Technikerstra{\ss}e 25, 6020 Innsbruck, Austria
\and School of Physics, University of Sydney, NSW 2006, Australia \and Istituto di Radioastronomia - INAF, via Gobetti 101, 40129 Bologna, Italy 
 \and Laboratoire Cassiop\'ee, CNRS, Observatoire de la C\^ote d'Azur, BP4229, 
06304 Nice Cedex 4, France}
\date{Received~28 February 2006; accepted~29 June 2006}  

\abstract
{We present the analysis of our 13 and 22 cm ATCA observations of the
central $\sim$18$\times$15 ${\rm arcmin}^2$ region of the merging
galaxy cluster A3921 ($z$=0.094). We investigated the effects of the
major merger between two sub-clusters on the star formation (SF) and
radio emission properties of the confirmed cluster members. The origin
of SF and the nature of radio emission in cluster galaxies was
investigated by comparing their radio, optical and X-ray
properties. We also compared the radio source counts and the
percentage of detected radio galaxies with literature data. We
detected 17 radio sources above the flux density limit of 0.25
mJy/beam in the central field of A3921, among which 7 are cluster
members. 9 galaxies with star-forming optical spectra were observed in
the collision region of the merging sub-clusters. They were not
detected at radio wavelengths, giving upper limits for their star
formation rate significantly lower than those typically found in
late-type, field galaxies. Most of these star-forming objects are
therefore really located in the high density part of the cluster, and
they are not infalling field objects seen in projection at the cluster
centre. Their SF episode is probably related to the cluster collision
that we observe in its very central phase. None of the galaxies with
post-starburst optical spectra was detected down our 2$\sigma$ flux
density limit, confirming that they are post-starburst and not dusty
star-forming objects.  We finally detected a narrow-angle tail (NAT)
source associated with the second brightest cluster galaxy (BG2),
whose diffuse component is a partly detached pair of tails from an
earlier period of activity of the BG2 galaxy.
\keywords{Galaxies: clusters: general -- Galaxies: clusters: individual: 
Abell 3921 -- Radio continuum: galaxies}}
\maketitle   

\section{Introduction}

In the past 25 years, optical observations of clusters have revealed
an evolution in galaxy properties with redshift. The fraction of
star-forming and post-star-forming cluster objects significantly
increases with $z$, ranging from $\sim$1--2\% in the local Universe
(Dressler 1987) to $\geq$30\% at $z{\sim}$0.3--0.5 (Dressler et
al. 1999). In 1978, Butcher \& Oemler reported a strong evolution of
the mean galaxy colour with redshift, from red to blue with increasing
$z$. A debate on the physical origin of this effect began: the
observed change of galaxy colours with redshift could be simply due to
a passive evolution of infalling field galaxies, or it could be
strongly affected by environmental effects. In 1983 Dressler
\& Gunn pointed out for the first time that in fact the blue colour of
the population detected by Butcher \& Oemler was the result of star
formation (SF) activity. Since then, many different studies have tried
to explain the origin of the observed evolution in the SF history of
cluster galaxies. The different physical mechanisms that have been
proposed may either trigger or weaken SF within clusters
(e.g. Dressler \& Gunn 1983; Evrard 1991; Bekki 1999; Fujita et
al. 1999).

The hierarchical model of structure formation predicts the formation
and evolution of galaxy clusters through mergers of less massive
systems. Due to the large energies involved in cluster-cluster
collisions, a merging event can enhance the efficiency of the physical
mechanisms responsible for the evolution of the galaxy star formation
rate (SFR). The first observational evidence for a correlation between
cluster mergers and SF activity came from the optical analysis of the
Coma cluster, where an excess of star-forming and post-star-forming
galaxies was observed in the collision region between the main cluster
and a group of galaxies (Caldwell et al. 1993). While in several other
non-relaxed clusters the collision between two or more clumps seems to
have increased the SFR of the galaxies (e.g. Gavazzi et al. 2003),
other analyses show the opposite trend (e.g. Baldi et
al. 2001). Therefore, the net role played by the merging event on SF
still has to be fully understood. To answer this question,
multi-wavelength (optical, X-ray, radio and/or IR) observations of
galaxy clusters are essential, as they allow us to determine in detail
both the dynamical state of the observed systems and the SF properties
of cluster members (e.g. Dwarakanath \& Owen 1999; Owen et al. 1999;
Venturi et al. 2000, 2001 and 2002; Duc et al. 2002; Miller, Owen \&
Hill 2003; Rizza et al. 2003; Giacintucci et al. 2004; Owen et
al. 2005a and 2005b).

Abell 3921 is a R=2, BM II Abell cluster at $z=0.094$ (Katgert et
al. 1996, 1998). Its perturbed morphology, first revealed by ROSAT and
Ginga observations (Arnaud et al. 1996), has been recently confirmed
by new optical (multi-object spectroscopy and VRI deep imaging,
Ferrari et al. 2005, F05 in the following) and X-ray analyses (XMM
observations, Belsole et al. 2005, B05 in the following). Two dominant
clumps of galaxies with a distorted morphology and a mass ratio of
$\sim$5 have been detected: a main cluster centred on the Brightest
Cluster Galaxy (BCG) (A3921-A), and an NW sub-cluster (A3921-B)
hosting the second brightest cluster member.  The comparison of the
optical and X-ray properties of A3921 (i.e. dynamical and kinematic
properties of the cluster, optical and X-ray morphology, features in
the ICM density and temperature maps) suggests that A3921-B is
probably tangentially traversing the main cluster along the SW/NE
direction. The two sub-clusters are observed in the central phase of
their merging process ($t_{coll}\sim\pm$ 0.3 Gyr), with a collision
axis approximately on the plane of the sky (F05; B05). A comparison of
the metallicity distribution in A3921 with numerical simulations
refined the reconstructed merging scenario, suggesting that we are
observing the very central phase of the clusters' interaction, just
before the first core-core encounter (Kapferer et al. 2006).

From spectral features of galaxies belonging to A3921, the SF
properties of confirmed cluster members have also been studied
(F05). Substantial fractions of both emission-line and
post-star-forming objects (so called k+a's) have been detected,
comparable to those measured at intermediate redshifts. Most of the
k+a galaxies in A3921 have fainter magnitudes than the
post-star-forming objects detected in higher redshift clusters
(``downsizing effect'', e.g. Poggianti et al. 2004). Since
emission-line galaxies in A3921 share neither the kinematics nor the
projected distribution of the passive cluster members, and they are
mostly concentrated in the region between the two sub-clusters, F05
concluded that the ongoing merger may have triggered a SF episode in
the galaxies located inside or around the collision region, where a
hot compression bar has been detected in the ICM temperature
distribution (B05). The on-going merger could also be responsible for
the high fraction of star-forming and post-star-forming objects in
this low redshift cluster.

These results motivated new high sensitivity radio observations of
A3921 with the Australia Telescope Compact Array (ATCA), in order to
shed light on the possible connection between the detected merger and
the SF and AGN activity in cluster galaxies.  This paper deals with
the analysis of the radio emission in the central region of the
cluster, where we detected the major merger between the main
sub-clusters A3921-A and A3921-B.  Sect.~\ref{DR} briefly describes
the observations and the data reduction. In Sects.~\ref{RLF} and
\ref{SF-AGN} we analyse the radio emission from the A3921 central
region and from the confirmed cluster members. The nature of the
peculiar radio source associated with the second brightest cluster
galaxy is discussed in Sect.~\ref{J2249-6420}. The main results and
their interpretation are summarized in Sect.~\ref{disc}.

As in F05 and B05, throughout the paper we assume ${\rm H}_0$=75
km~${\rm s}^{-1}~{\rm Mpc}^{-1}$, $\Omega_{\rm m} = 0.3$,
$\Omega_{\Lambda} = 0.7$. In this cosmology 1~arcmin corresponds to
$\sim$0.097~\h~Mpc.

\section{Observations and data reduction}\label{DR}

\begin{table}
\begin{center}   
\begin{tabular}{lcc}  
\hline
\hline
Configuration & \multicolumn{2}{c}{Baselines (metres)}\\ 
& Min & Max \\
\hline
6A & 337 & 5939\\
6C & 153 & 6000\\
750C & 46 & 5020\\
1.5D & 107 & 4439\\
\hline
\end{tabular}
\caption{Minimum and maximum spacings between antennas in the East-West array configurations used for our ATCA 
observations. }
\label{tab:configurations}
\end{center}
\end{table}

\begin{figure*}
\begin{center}
\resizebox{13cm}{!}  
{\includegraphics{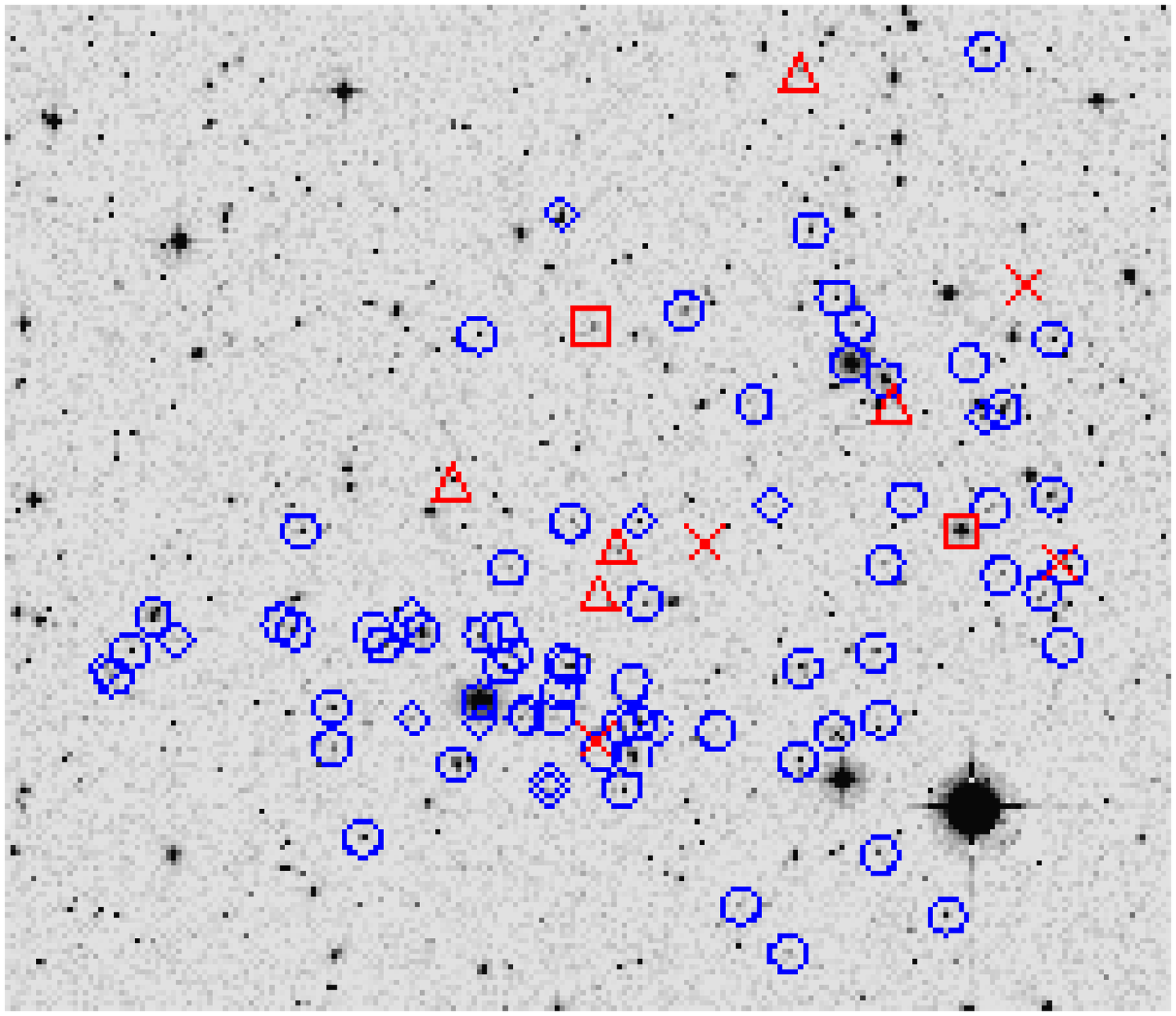}}
\resizebox{13cm}{!}  
{\includegraphics{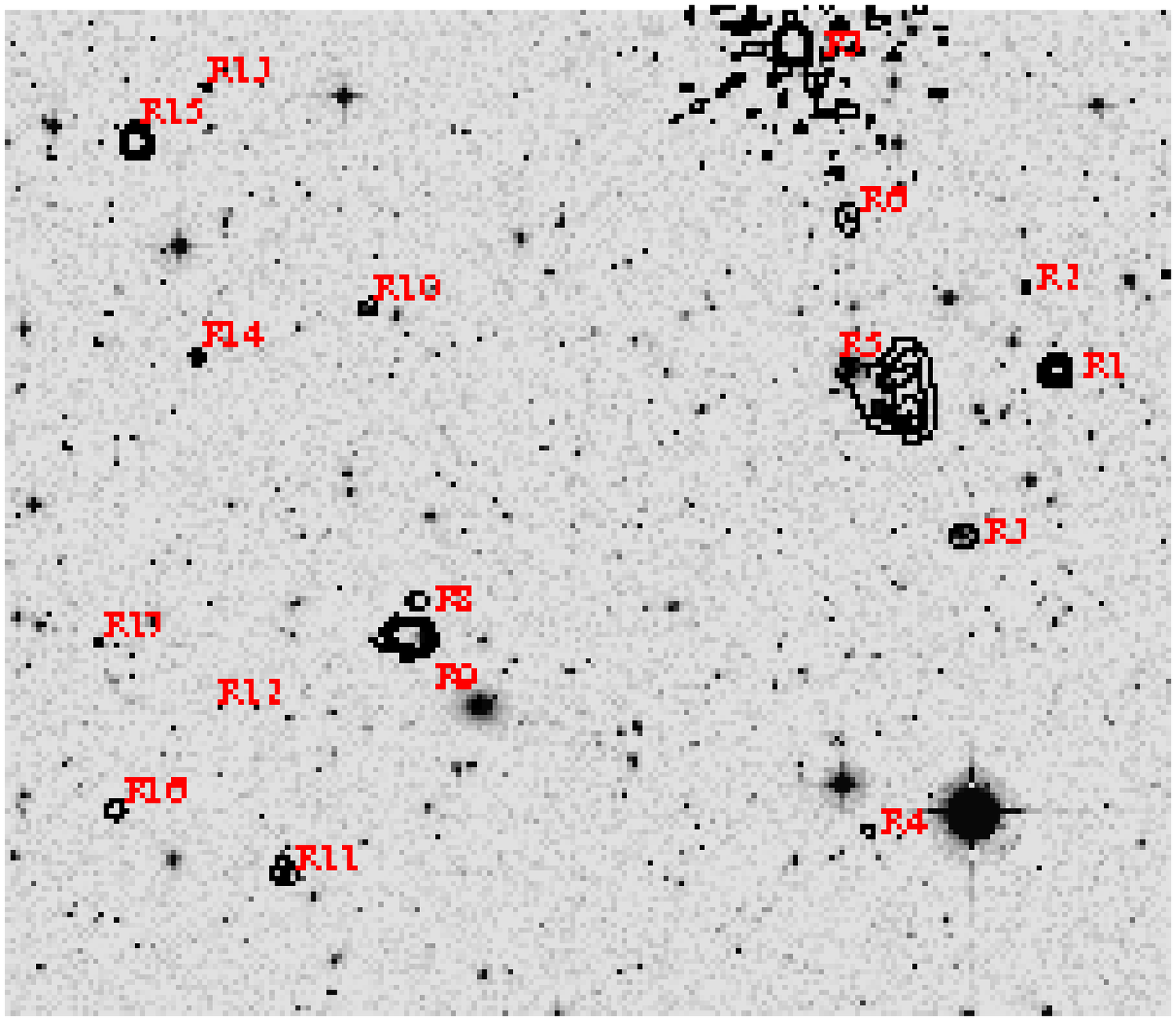}}
\hfill
\parbox[b]{18cm}{
\caption{{\bf Top:} R-band image of the central field 
($\sim$18$\times$15 ${\rm arcmin}^2$) of A3921. Overlaid symbols
correspond to A3921 confirmed cluster members. Based on their spectral
properties, galaxies were classified as old passively evolving
ellipticals (circles), e(b)'s (squares), e(a)'s (triangles), e(c)'s
(crosses), k+a's (diamonds) (see text and F05 for more details). {\bf
Bottom:} 1.344~GHz radio contours are overlaid on the R-band central
field of A3921. Contour levels start at 0.245 mJy/beam
(5$\times$r.m.s. of the 22~cm map), with linear steps of 1.000
mJy/beam up to 4.245 mJy/beam. The radio ID of each source is also
shown (see Table~\ref{tab:5s}).}
\label{fig:FOC_radio_ottico}}
\end{center}
\end{figure*}

We observed the central $\sim$18$\times$15 ${\rm arcmin}^2$ 
($\sim$1.8$\times$1.5 ${\rm Mpc}^2$) field of A3921 covered by our 
previous optical observations with the Australia Telescope Compact 
Array (ATCA) at 22 cm ($\nu$=1.344 GHz) and 13 cm ($\nu$=2.368 GHz) 
simultaneously. The ATCA comprises five 22 m antennas placed on a 3 
km railway track, running east-west, and an isolated sixth antenna, 
located 3 km further to the west. There are several stations on 
which the antennas can be placed, thus allowing a selection of 
extended or compact configurations with a range of different 
spacings.  The configurations used here, which together gave 
reasonably uniform uv coverage, are given in 
Table~\ref{tab:configurations}, where the minimum and maximum 
spacings are indicated.  Observations were performed in continuum 
mode with a bandwidth of 128 MHz in May, July, November and December 
2004. A uniform sensitivity radio map of the central 18$\times$15 
${\rm arcmin}^2$ of A3921 was obtained by using the mosaicing 
facility of the ATCA. The central cluster field was covered with 
eight overlapping pointings separated by 10 arcmin (i.e. half the 
primary beam width at 13 cm). To obtain good hour angle coverage, 
the eight pointings were observed in sequence every 160 seconds. The 
cycle, including calibration, was repeated for 12 hours for each 
observing run. B1934$-$638 and B2205$-$636 were used as primary and 
secondary calibrators. The largest structure detectable by these 
observations corresponds to $\sim$16 and 10~arcmin at 22 and 13 cm 
respectively.

Data reduction was carried out following the standard procedure
(calibration, Fourier inversion, clean and restore) with the {\it
MIRIAD\/} package. We reduced and imaged each pointing separately. The
eight cleaned maps at 22 and 13 cm were finally mosaiced
(i.e. linearly combined) with the {\it MIRIAD\/} task {\it LINMOS},
using the {\it taper\/} option of this program, which corrects for
primary beam attenuation avoiding excessive noise amplification at the
edge of the mosaic field (Sault, Staveley-Smith
\& Brouw 1996).  The noise levels in the final maps are given 
in Table~\ref{tab:mosaic}. We also obtained a spectral index
(${\alpha}_{13}^{22}$) map by combining the images at 1.344~GHz and
2.368~GHz with the {\it AIPS} task {\it COMB}. For this purpose, maps
at the two frequencies were made with the same beam
(i.e. $12{\times}12$ arcsec) and the same uv-range (i.e.  0.354--30
k$\lambda$). The output image was blanked whenever the intensity of
the corresponding pixel was lower than twice the map noise in either
image.

\section{Radio emission from the central field of A3921}\label{RLF}

\subsection{The sample}\label{sample}

Within the central $\sim$18$\times$15 ${\rm arcmin}^2$ of A3921 we
found 17 radio sources above a peak flux density limit of 5 times the
r.m.s. noise in the 1.344 GHz map. The dynamic range is of the order
of 1000:1, so the local noise level around strong sources may be
higher than that given in Table~\ref{tab:mosaic}.

The source list is reported in Table~\ref{tab:5s}, where we give:

\begin{itemize}

\item[-] {\it Columns (1) \& (2):} radio ID and name of detected
sources.  We used the prefix ``FHF2006'' (i.e. initials of the first
three authors and year of publication) for the names of sources not
present in any previous radio survey.

\item[-] {\it Columns (3) \& (4):} right ascension and declination 
(J2000.0) with errors.

\item[-] {\it Columns (5), (6) \& (7):} integrated intensity at 22 
and 13 cm and spectral index with errors\footnote{The radio spectral
index $\alpha$ is defined such as $S_{\nu}{\propto}{\nu}^{\alpha}$.}.

\item[-] {\it Column (8):} radio morphology (``resolved'' or 
``unresolved'').

\item[-] {\it Column (9):} measured angular size of the 
source, if resolved.

\end{itemize}

\begin{table}
\begin{center}   
\begin{tabular}{ccc}  
\hline
\hline
$\nu$ & RMS Noise & $b$ \\
GHz & mJy/beam & arcsec \\
\hline
1.344 & 0.049 & 12.0 \\
2.368 & 0.052 & 8.0 \\
2.368 & 0.053 & 12.0 \\

\hline
\end{tabular}
\caption{RMS noise in the 1.344 and 2.368 GHz images. The FWHM 
of the restoring beam is shown in the third column.}
\label{tab:mosaic}
\end{center}
\end{table}

\begin{table*}
\begin{center}   
\begin{tabular}{llccccccc}  
\hline
\hline
\multicolumn{1}{c}{ID$_{\rm R}$} & {Source name} & {${\rm RA}_{\rm 
J2000}$(err)} & {${\rm Dec}_{\rm J2000}$(err)} & {${\rm S}_{22cm}$(err)} & {${\rm S}_{13cm}$(err)} & {${\alpha}_{13}^{22}$(err)} & {Mor.} & {Size} \\ 
\# & & $^{h~m~s}$ $^{(s)}$ & $^{\circ~'~''}$ $ ^{('')}$ & mJy & mJy & & & $ ^{''~{\times}~''}$ \\ 
 & & & & & & & & deg \\
\hline
R1 & SUMSS J224834$-$642037 & 22 48 35.24(0.03) & $-$64 20 37.3(0.5) 
& 
11.79(0.14) & 8.11(0.14) & $-$0.66(0.04) & u & --- \\
 & & & & & & & & --- \\
R2 & FHF2006 J224839$-$641921 & 22 48 39.74(0.13) & $-$64 19 
21.7(1.1) 
& 0.36(0.08) & 0.31(0.11) & $-$0.3: & u & --- \\
 & & & & & & & & --- \\
R3 & FHF2006 J224848$-$642311 & 22 48 48.74(0.06) & $-$64 23 
11.4(0.6) 
& 1.80(0.05) & 1.32(0.05) & $-$0.55(0.08) & r & 16$\times$14, \\
  & & & & & & & & p.a.=102 \\
R4 & FHF2006 J224902$-$642743 & 22 49 02.60(0.04) & $-$64 27 
43.9(0.5) 
& 0.37(0.06) & 0.21(0.07) & $-$1.0: & u & --- \\
 & & & & & & & & --- \\
R5 & SUMSS J224857$-$642059 & 22 49 04.63(0.03) & $-$64 20 36.3(0.5) 
& 
43.80(0.44) & 13.95(0.15) & $-$2.02(0.03) & r & 105$\times$99, \\
 & & & & & & & & p.a.=0 \\
R6 & FHF2006 J224904$-$641816 & 22 49 04.98(0.04) & $-$64 18 
16.3(0.5) 
& 1.90(0.07) & 1.29(0.07) & $-$0.68(0.11) & r & 16$\times$12, \\
 & & & & & & & & p.a.=16 \\
R7 & PMN J2249$-$6415 & 22 49 12.73(0.03) & $-$64 15 34.0(0.5) & 
133.20(1.35) & 81.02(0.89) & $-$0.88(0.03) & u & --- \\
 & & & & & & & & --- \\
R8 & FHF2006 J225007$-$642408 & 22 50 07.00(0.05) & $-$64 24 
08.9(0.5) 
& 0.88(0.09) & 0.59(0.43) & $-$0.7: & u & --- \\
 & & & & & & & & --- \\
R9 & SUMSS J225007$-$642439  & 22 50 07.72(0.03) & $-$64 24 
40.9(0.5) 
& 63.82(0.64) & 42.55(0.43) & $-$0.72(0.03) & r & 21$\times$12, \\
 & & & & & & & & p.a.=81 \\
R10 & FHF2006 J225013$-$641935 & 22 50 13.64(0.07) & $-$64 19 
35.3(0.6) 
& 0.82(0.05) & 0.33(0.05) & $-$1.61(0.29) & r & 15$\times$12, \\
 & & & & & & & & p.a.=124 \\
R11 & FHF2006 J225026$-$642819 & 22 50 26.41(0.04) & $-$64 28 
19.8(0.5) 
& 2.18(0.05) & 1.20(0.05) & $-$1.06(0.08) & r & 14$\times$13, \\
 & & & & & & & & p.a.=27 \\
R12 & FHF2006 J225035$-$642542 & 22 50 35.61(0.15) & $-$64 25 
42.8(1.3) & 0.25(0.07) & 0.2: & --- & u & --- \\
 & & & & & & & & --- \\
R13 & FHF2006 J225036$-$641608 & 22 50 36.52(0.11) & $-$64 16 
08.4(0.7) & 0.33(0.09) & 0.3: & --- & u & --- \\
 & & & & & & & & ---\\
R14 & FHF2006 J225038$-$642020 & 22 50 38.24(0.11) & $-$64 20 
20.1(1.2) 
& 0.80(0.05) & 0.64(0.05) & $-$0.40(0.19) & r & 24$\times$14, \\
 & & & & & & & & p.a.=7 \\
R15 & SUMSS J225046$-$641658 & 22 50 46.60(0.03) & $-$64 16 
59.4(0.5) 
& 21.04(0.23) & 12.89(0.37) & $-$0.87(0.05) & u & --- \\
 & & & & & & & & --- \\
R16 & FHF2006 J225050$-$642720 & 22 50 50.59(0.05) & $-$64 27 
20.0(0.6) & 0.82(0.08) & 0.4: & --- & u & --- \\
 & & & & & & & & ---\\
R17 & FHF2006 J225052$-$642443 & 22 50 52.57(0.11) & $-$64 24 
43.8(0.9) & 0.30(0.07) & 0.4: & --- & u & --- \\ 
 & & & & & & & & ---\\
\hline
\end{tabular}
\caption{List of $5\sigma$ sources in the central field 
($\sim$18$\times$15 ${\rm arcmin}^2$) of A3921.  Their position,
integrated intensity at 22 and 13 cm, mean spectral index and
morphology (``r''=resolved, ``u''=unresolved) are given.  Errors are
indicated in brackets. The colon indicates a highly uncertain
value. Spectral index is not given for highly uncertain 13 cm
fluxes. For resolved sources, the measured size was obtained from a
Gaussian fit to the 22 cm image, except for R5 (SUMSS
J224857$-$642059; see text).}
\label{tab:5s}
\end{center}
\end{table*}

\noindent The total flux density and the position of the flux
density peak at 22 cm of the irregular resolved source R5 (SUMSS
J224857-642059) were calculated using the {\it AIPS} task {\it
TVSTAT}. Its size was estimated using the verb {\it TVDIST}. For all
the other regular (circular/elliptical) sources the position,
integrated intensity and measured size were obtained with a Gaussian
fit using the {\it AIPS} task {\it IMFIT}. The errors in Gaussian fits
were taken into account following Condon (1997). We fitted six free
parameters (amplitude $A$, central position ($x_0,y_0$), major and
minor axes ($\theta_M,\theta_m$), and position angle $\beta$). We
retained the fitted total flux density (${\rm S}_t$) only for resolved
sources. In the other cases the quoted total flux density corresponds
to the corrected amplitude, i.e. the fitted amplitude (${\rm S}_p$)
multiplied by the square root of the ratio of the fitted size of the
source (${\theta_M\theta_m}$) and the beam size
(${\theta_M^*\theta_m^*}$, eq. (29) of Condon 1997):

\begin{equation}
{\rm S}_t={\rm S}_p
\left(\frac{\theta_M\theta_m}{\theta_M^*\theta_m^*}\right)^{1/2}.
\label{eq:cond}
\end{equation}

\noindent Unresolved or barely resolved sources were identified at 22 
cm as those whose fitted size is within 1$\sigma$ of the map beam
size, and/or:

\begin{equation}
\frac{{\rm S}_{t}-\Delta{\rm S}_{t}}{{\rm S}_{p}+\Delta{\rm 
S}_{p}}<1,
\end{equation}

\begin{figure}
\begin{center}
\resizebox{8cm}{!}  
{\includegraphics{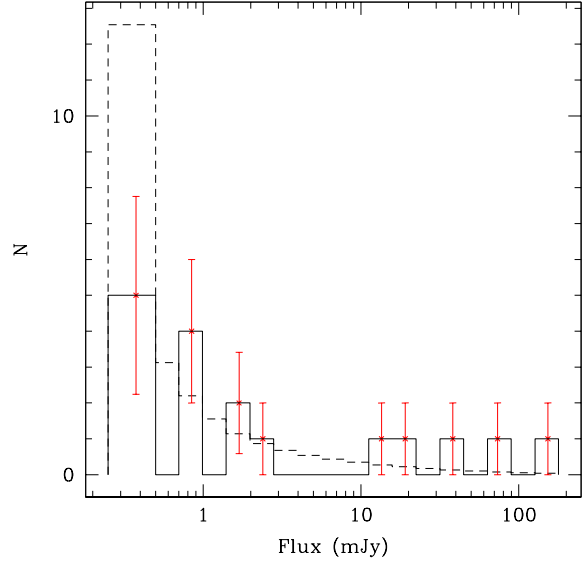}}
 \hfill  
\parbox[b]{8cm}{
\caption{Flux density distribution of the radio sources detected in A3921 
(solid line) and in the field (dashed line). The field counts, derived
from Hopkins et al. (2003), have been normalised to the sky area
considered in this paper (0.075 deg$^2$) 1$\sigma$ error bars
associated with our counts are also shown (assuming Poissonian 
noise).}
\label{fig:Ncounts}}  
\end{center}
\end{figure}

\noindent where ${\rm S}_{t}$ and ${\rm S}_{p}$ are the fitted
integrated and peak fluxes, and $\Delta{\rm S}_{t(p)}$ are the
associated errors:

\begin{equation}
\Delta{\rm S}_{t(p)}=\sqrt{{\rm \sigma_{Noise}}^2+(0.01{\rm 
S}_{t(p)})^2}.
\label{eq:errS}
\end{equation}

\noindent The errors in position ($\Delta{\alpha(\delta)}$) quoted 
in Table~\ref{tab:5s} were estimated as the quadratic combination of
the formal {\it IMFIT} errors and a nominal 0.5$''$ calibration
uncertainty. The errors in flux density were similarly obtained as the
quadratic combination of the r.m.s. noise of the radio map and the
formal uncertainty in the total flux density. This latter error was
assumed as in Eq.~\ref{eq:errS} for resolved sources (i.e. $0.01{\rm
S}_{t}$), and estimated through error propagation from
Eq.~\ref{eq:cond} for unresolved sources.

\subsection{Comparison with background counts}

As mentioned in the previous Sect., we counted 17 radio sources with
${\rm S}_{22cm}\geq$0.25 mJy in the central area (0.075 deg$^2$) of
A3921. We then compared this number of radio sources with field
counts, in order to test the possible presence of an excess of radio
emission in the region of the cluster hosting the main merger between
A3921-A and A3921-B. The sensitivity of the radio image analysed in
this paper is uniform to the 5$\sigma$ flux density limit (0.25
mJy/beam) due to the overlapping of the 8-pointings of the mosaic in
the central region of A3921.

Our results were compared with those of Hopkins et al. (2003). By
integrating their polynomial fit for source counts, 25 field radio
sources are expected down to the flux limit of our radio
observations. Assuming Poissonian noise for our counts, we found a
slight deficit of radio sources in our cluster field at $\sim2\sigma$
level. Fig.~\ref{fig:Ncounts} shows the radio flux density
distribution in A3921 and in the field (Hopkins et al.  2003). At
$\sim$2$\sigma$ level, there is either a possible excess of radio
sources in the field in the lowest flux density bin (0.25--0.50 mJy),
or the A3921 counts are incomplete at the faint end. Based on a
two-sided Kolmogorov-Smirnov test, the probability that the two
distributions are taken from the same parent population is 14\%, so
the test is not conclusive.

\subsection{Radio-optical identification}\label{ROid}

Using our VRI catalogue (WFI data, F05), we found optical
identifications for 14 of the 17 radio sources above 5$\sigma$.  They
are listed in Table~\ref{tab:ID}, which contains:

\begin{itemize}

\item[-] {\it Column (1):} radio ID based on Table~\ref{tab:5s}.

\item[-] {\it Column (2):} optical ID based on Table A of F05 (when available).

\item[-] {\it Columns (3) \& (4):} optical coordinates.

\item[-] {\it Column (5):} offset $\Delta_{r-o}$ between the radio and optical coordinates.

\item[-] {\it Column (6):} parameter 
$\mathcal{R}=\Delta_{r-o}/({\sigma_o}^2+{\sigma_r}^2)^{1/2}$,
where ${\sigma_r}$ (given in Table~\ref{tab:5s}) and ${\sigma_o}$ (0.1
arcsec) are the radio and optical position errors respectively.

\item[-] {\it Columns (7), (8) \& (9):} V, R and I-band magnitudes (WFI 
observations, F05).

\item[-] {\it Column (10):} redshift of the source (when available). 

\item[-] {\it Column (11):} Log of the absolute spectral 
luminosity at 1.344~GHz (W/Hz), estimated when the redshift is
available from $L=4\pi{D_L}^2{S_{1.344~GHz}}(1+z)^{-(1+\alpha)}$
(Petrosian \& Dickey 1973), where $\alpha$ is the spectral index of
the source.

\end{itemize}

\begin{figure*}
\resizebox{18cm}{!}{\includegraphics{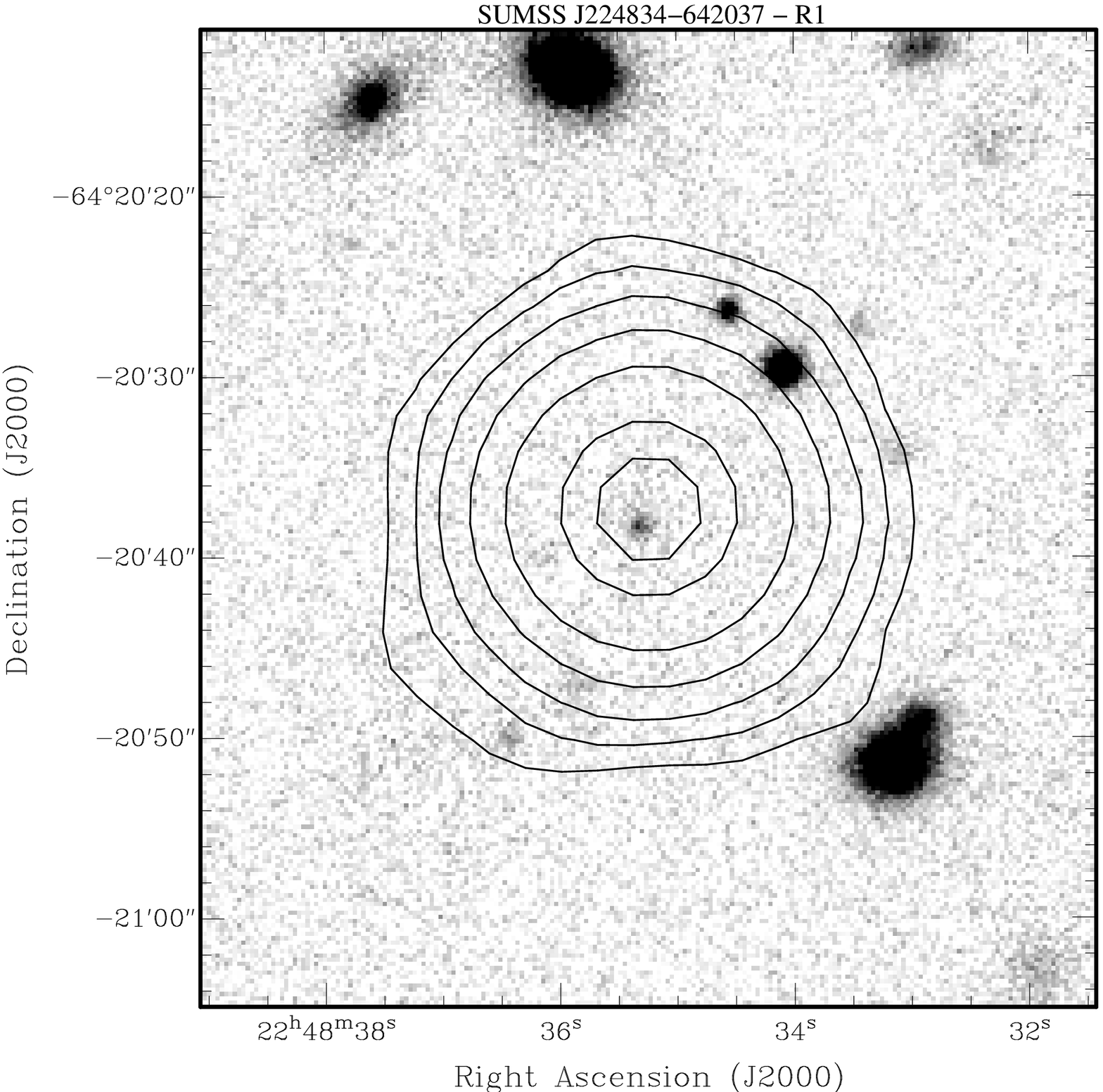}
\includegraphics{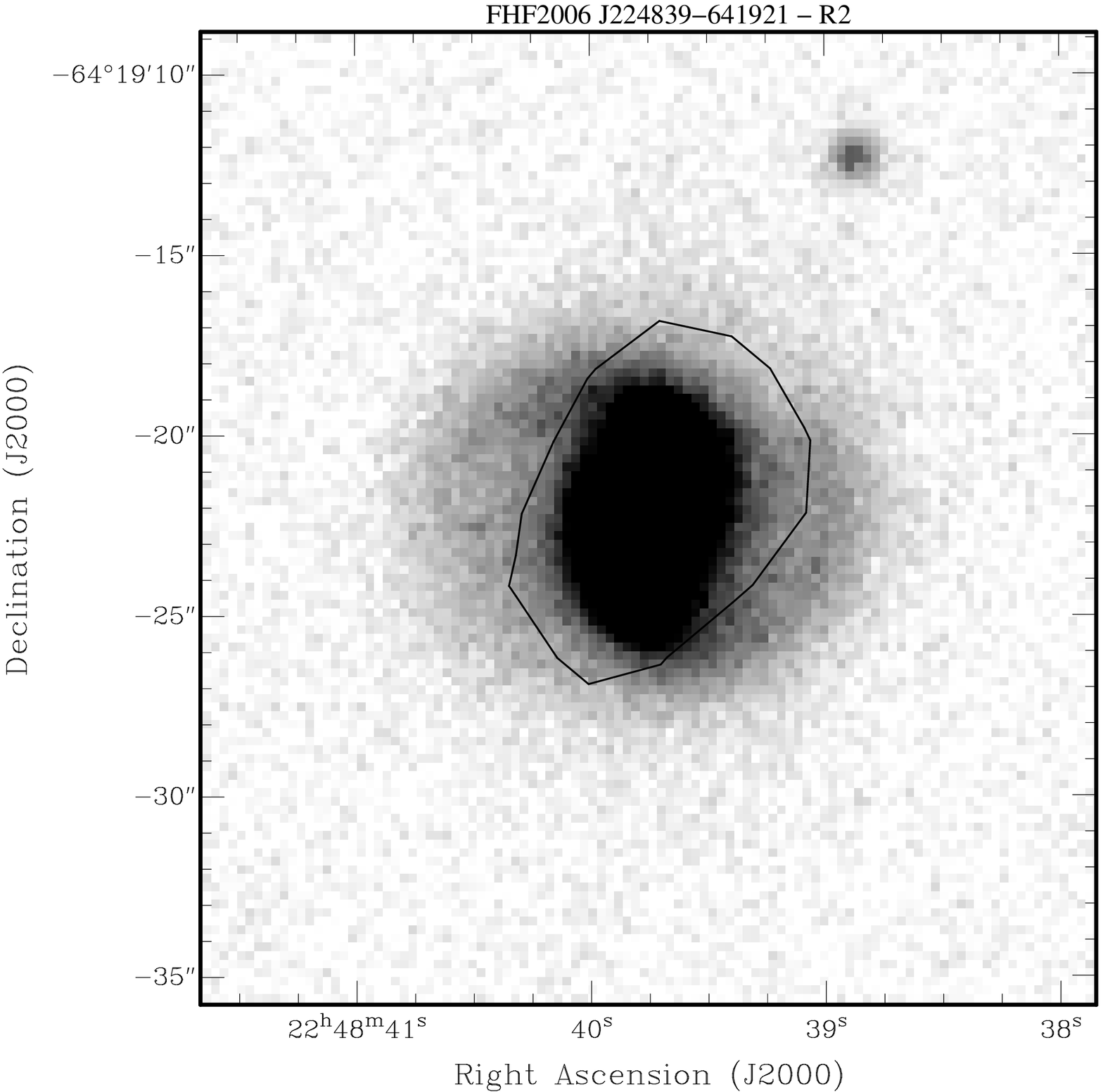}
\includegraphics{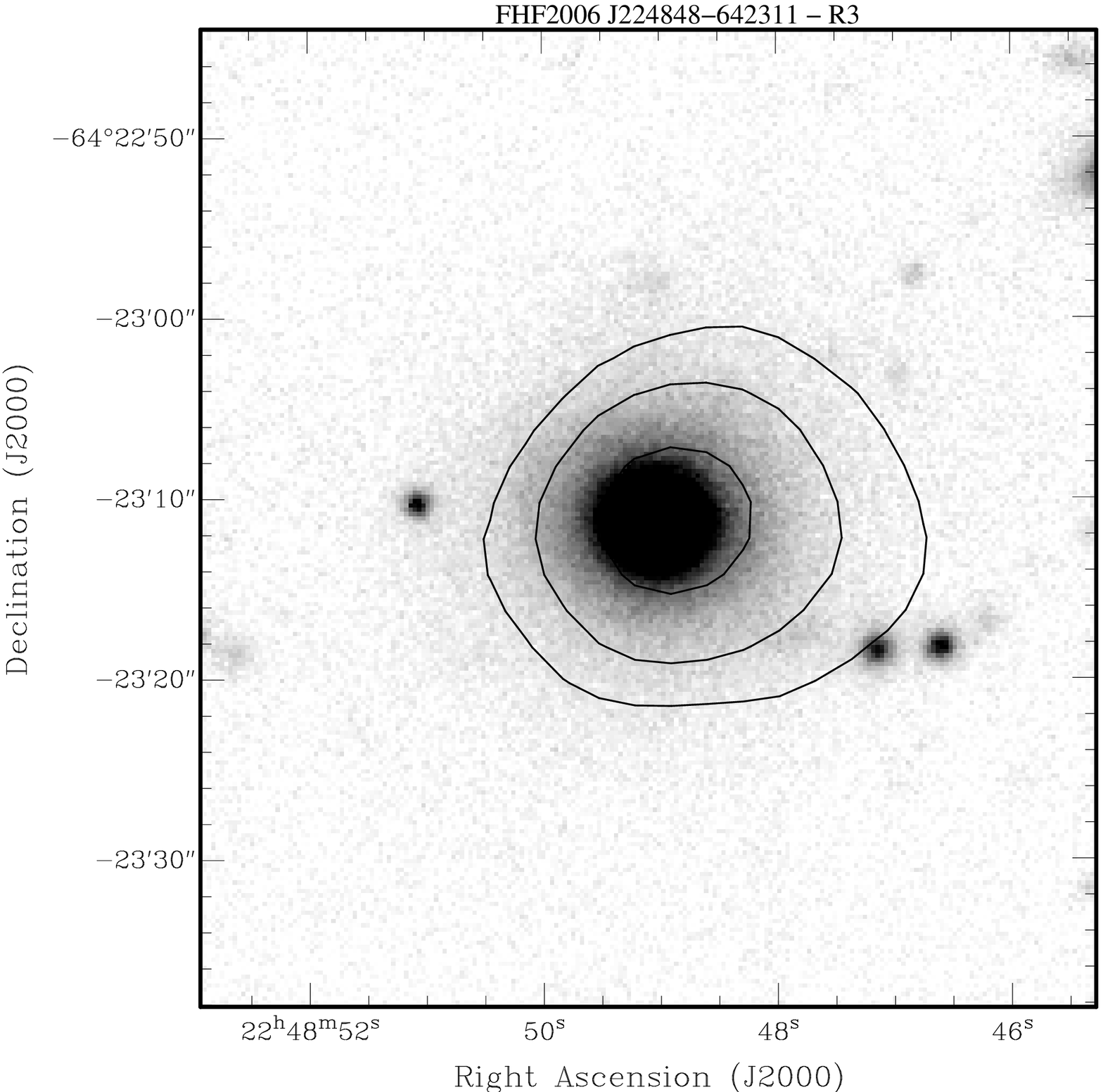}}
\resizebox{18cm}{!}{\includegraphics{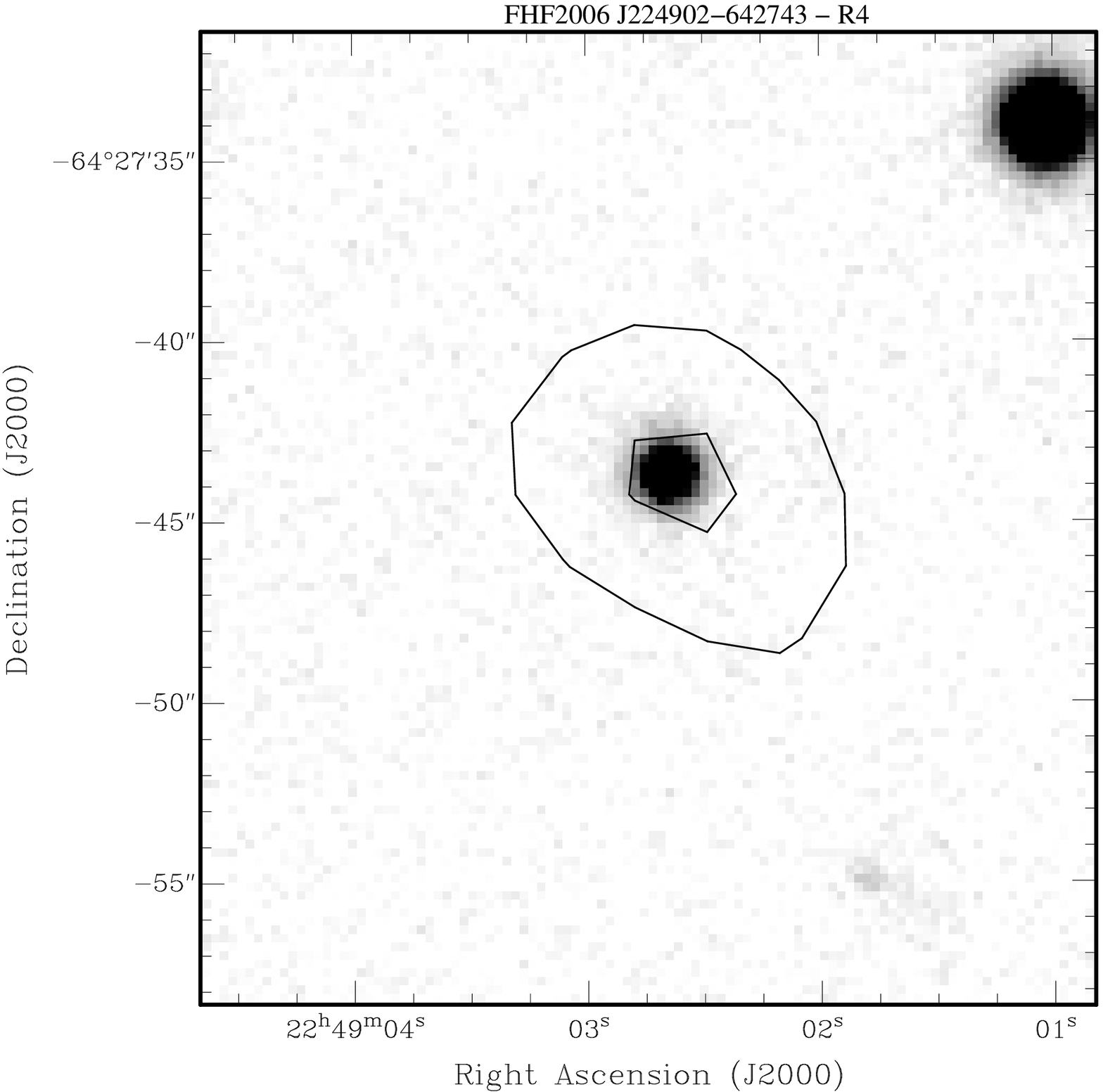}
\includegraphics{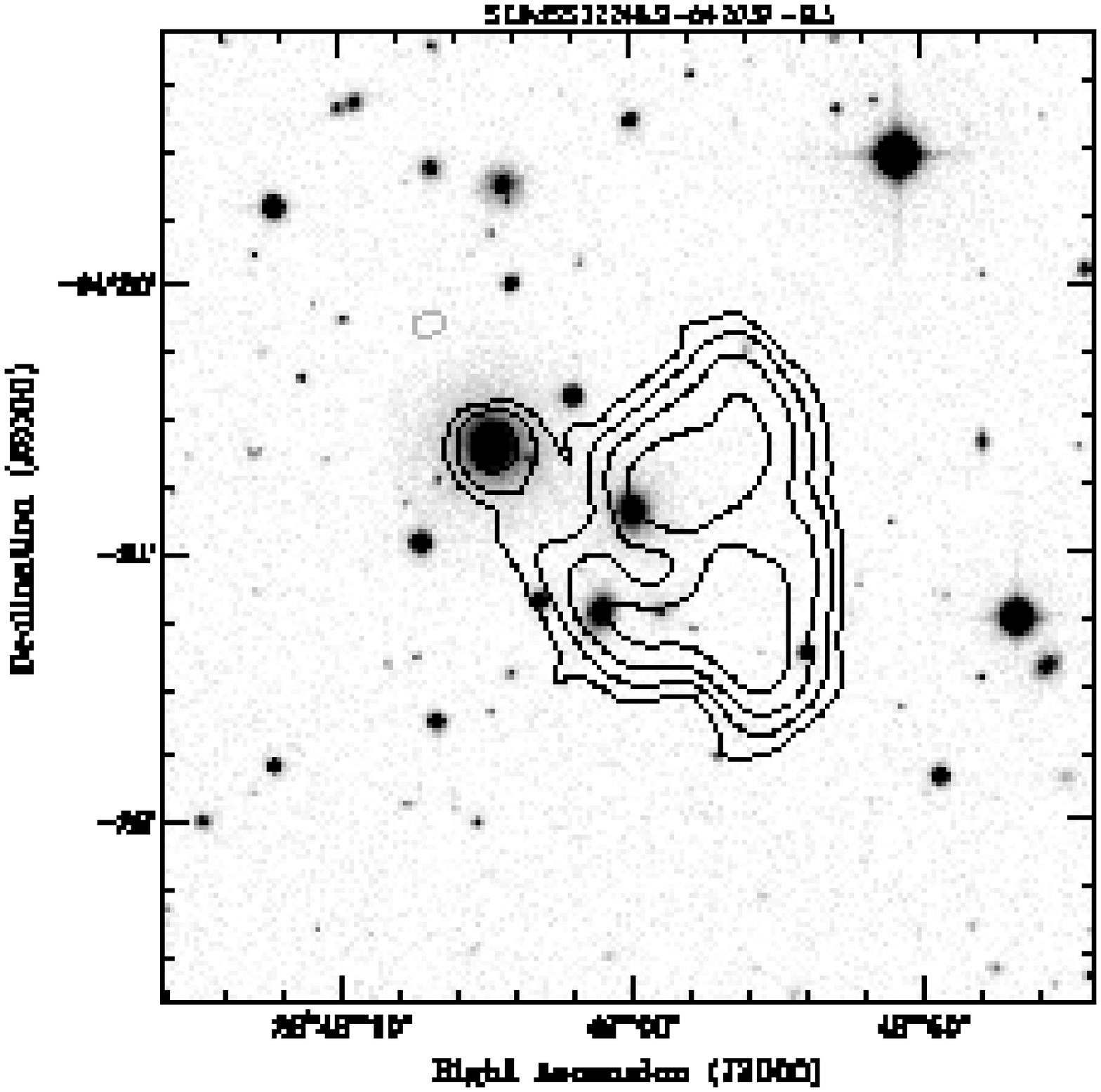}
\includegraphics{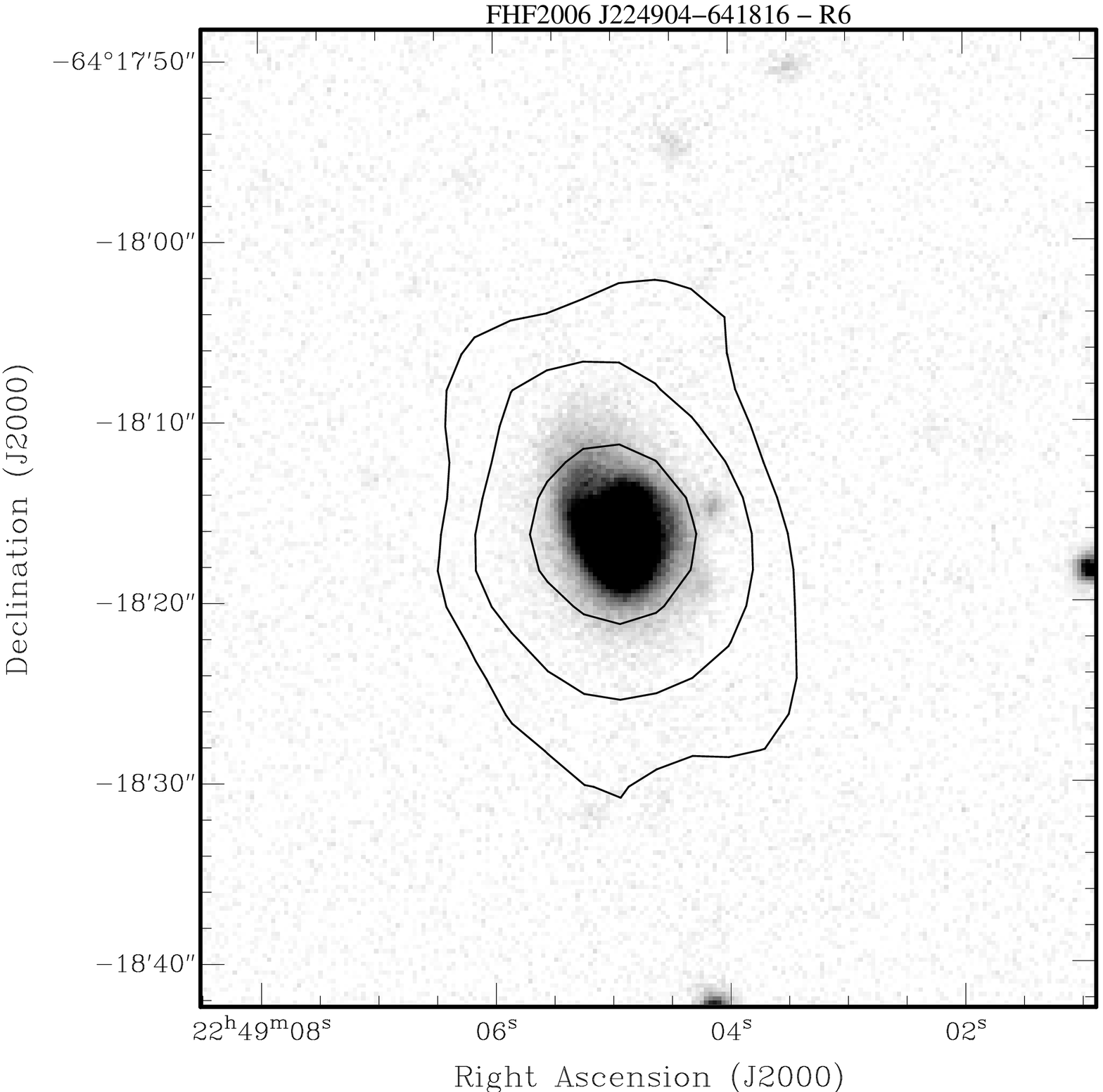}}
\resizebox{18cm}{!}{\includegraphics{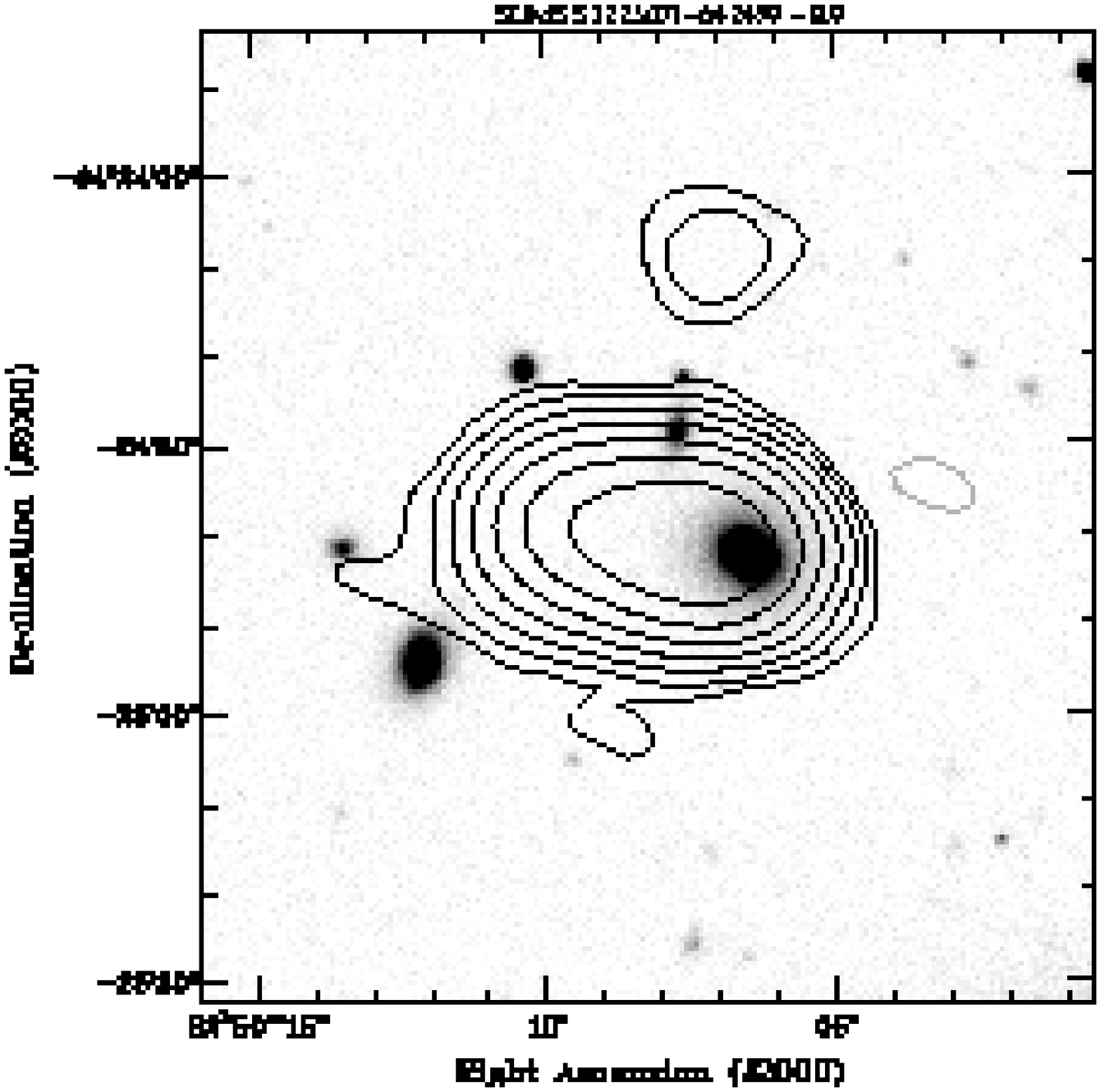}
\includegraphics{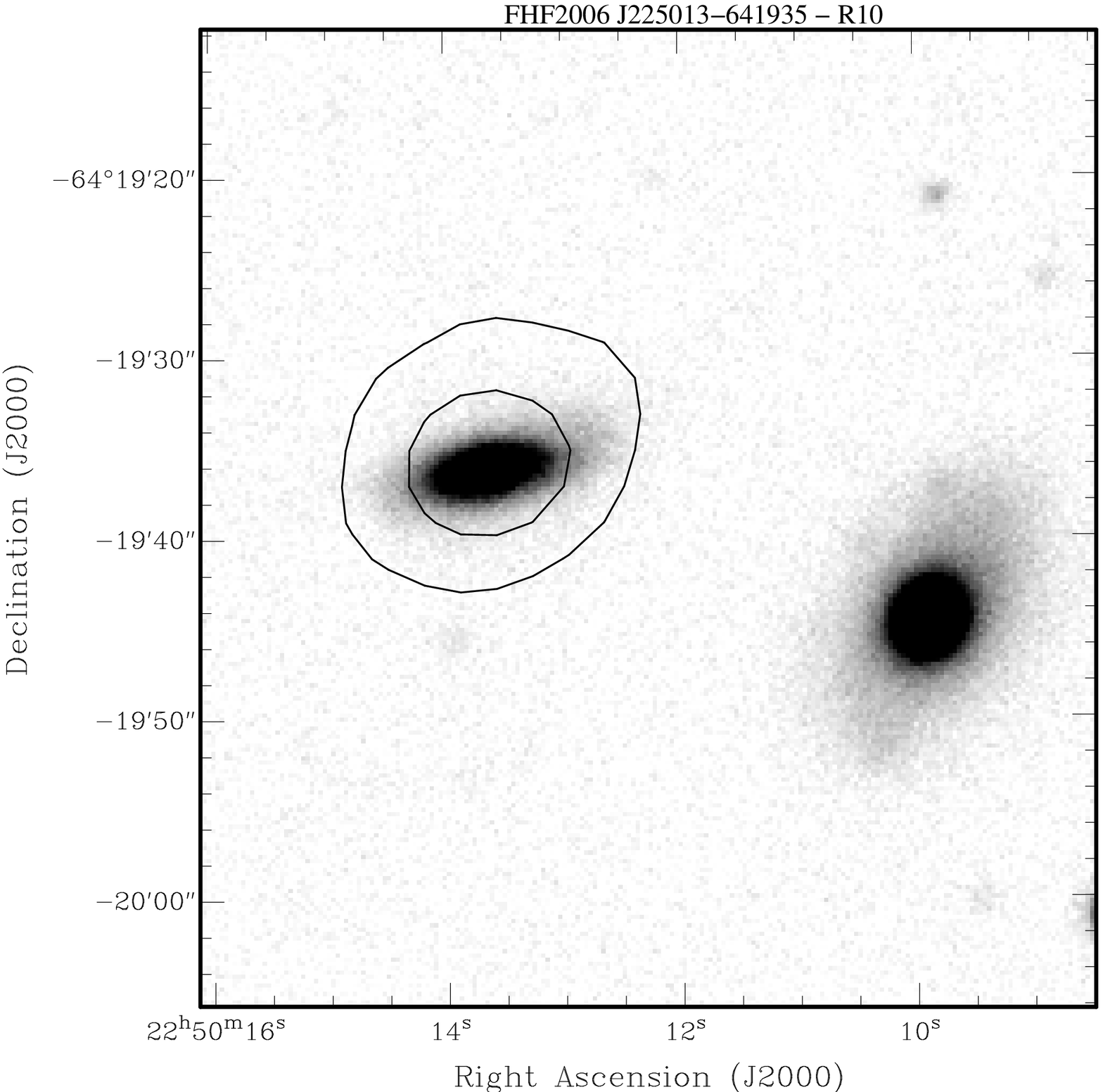}
\includegraphics{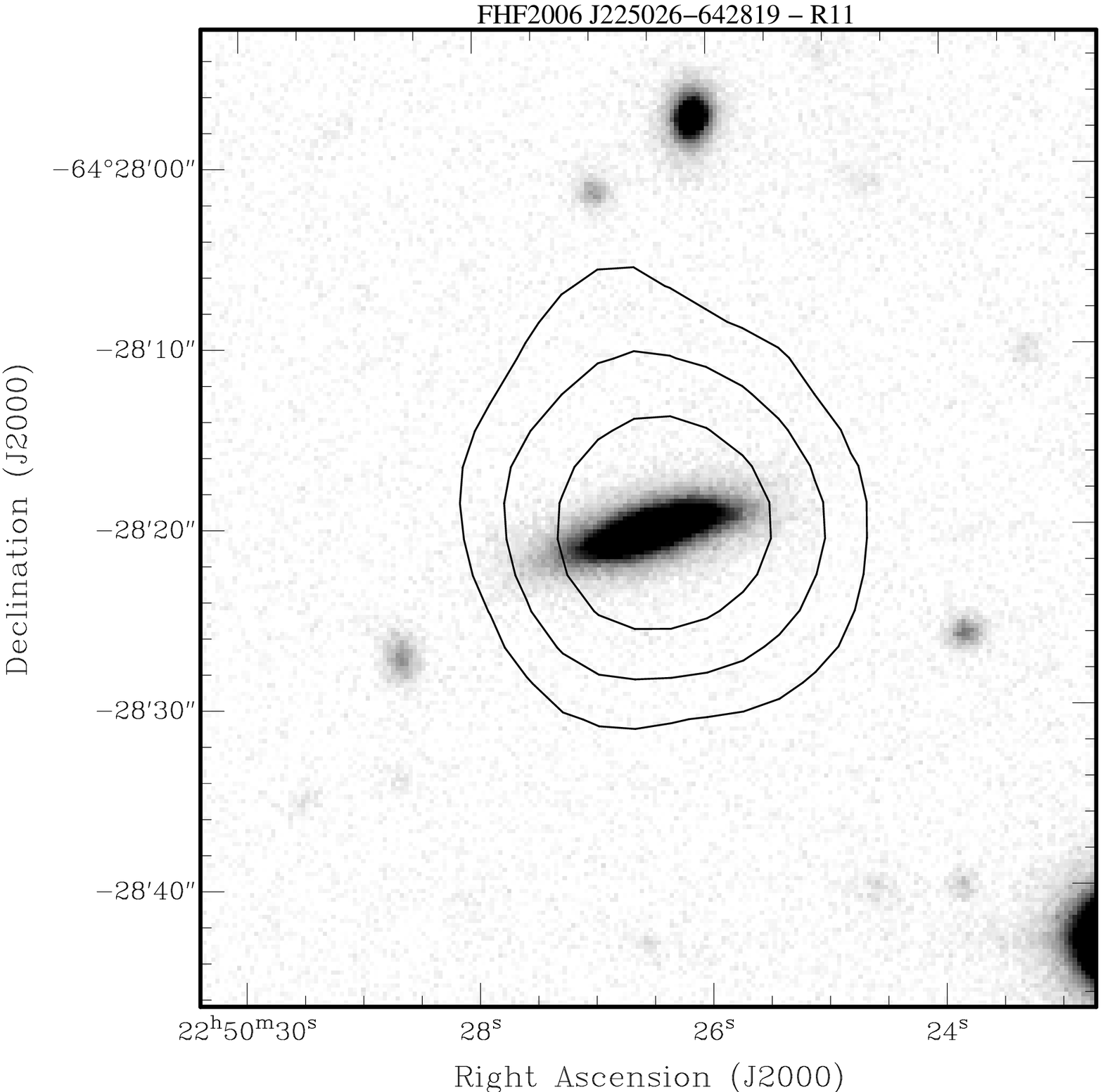}}
\hfill
\begin{center}
\parbox[b]{18cm}{
\caption{1.344~GHz radio contours of the radio sources in the A3921 
central field overlaid on the corresponding R-band image. When not 
specified, radio contours are as follow: 0.245 mJy/beam $\times$ 
($-$1, 1, 2, 4, 8, 16, 32, 64, ...). For R1 (SUMSS J224834$-$642037) 
radio contours are: 0.245 mJy/beam $\times$ ($-$1, 1, 2, 4, 8, 15, 
30, 40); for R4 (FHF2006 J224902$-$642743): 0.245 mJy/beam $\times$ 
($-$1, 1, 1.5). R3 (FHF2006 J224848$-$642311) and R5 (SUMSS 
J224857$-$642059) are associated with the third and the second 
brightest cluster galaxies (BG3 and BG2 in F05). }
\label{fig:ID}}
\end{center}
\end{figure*}

\setcounter{figure}{2}
\begin{figure*}
\resizebox{18cm}{!}{\includegraphics{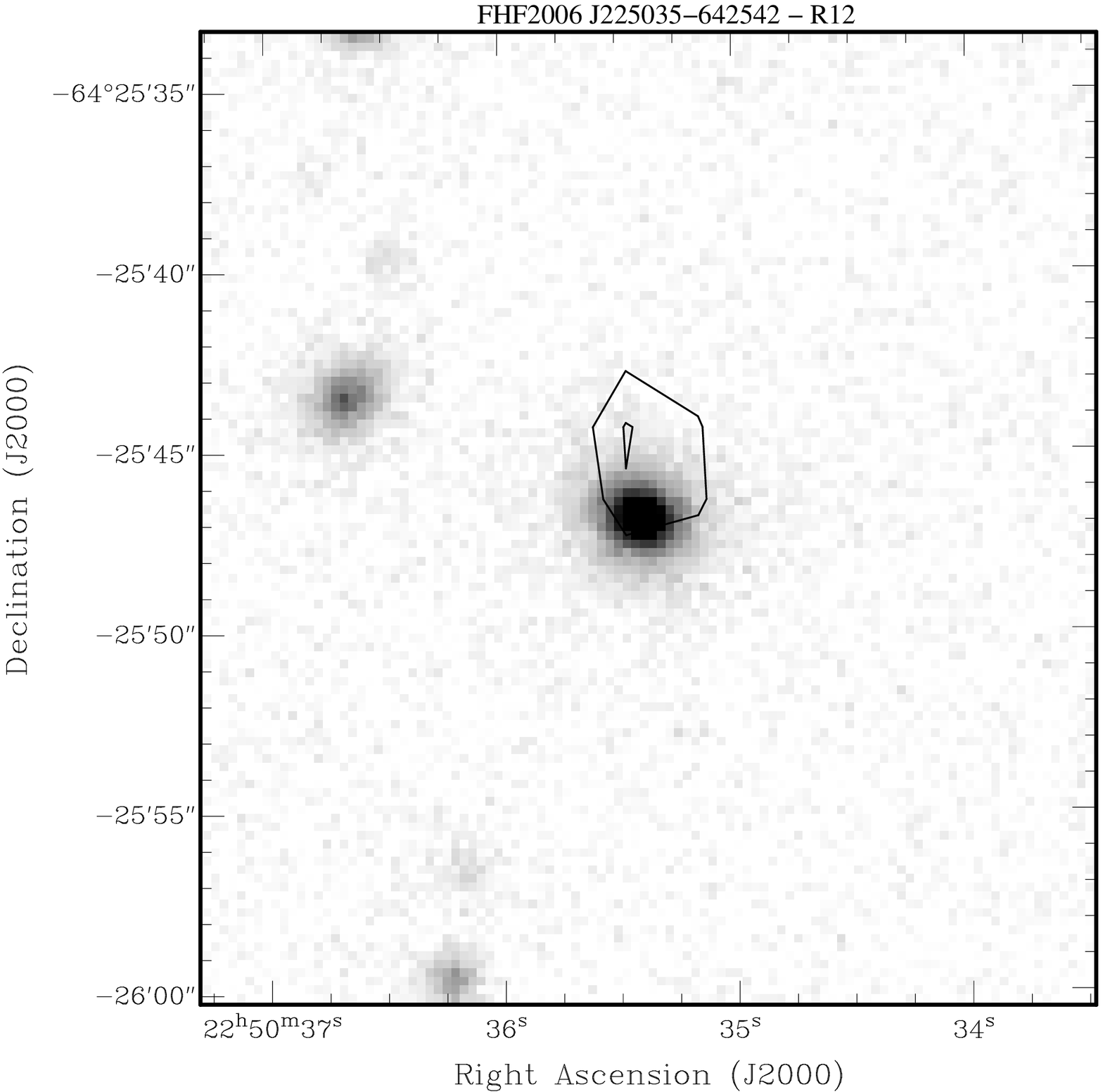}
\includegraphics{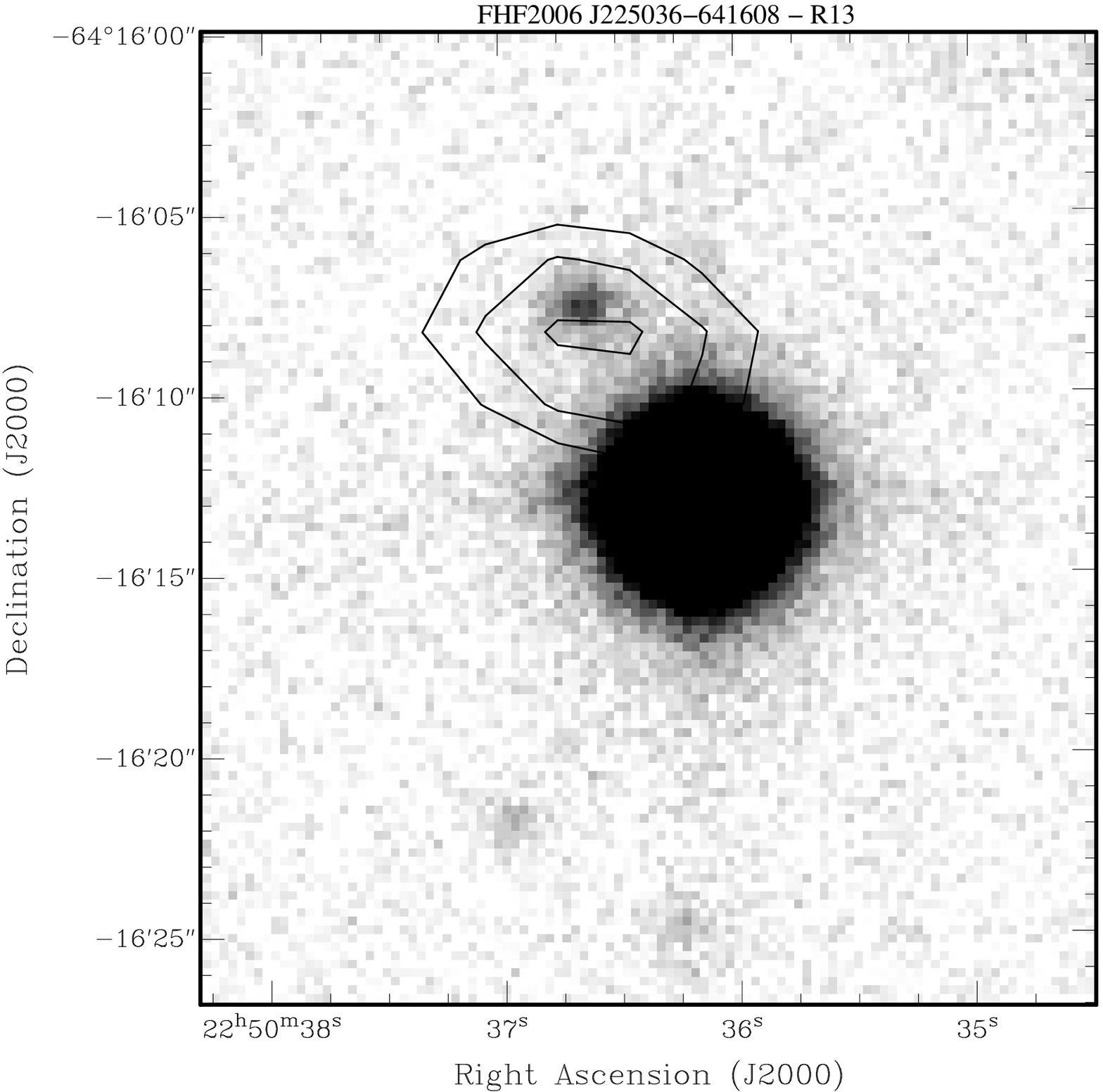}
\includegraphics{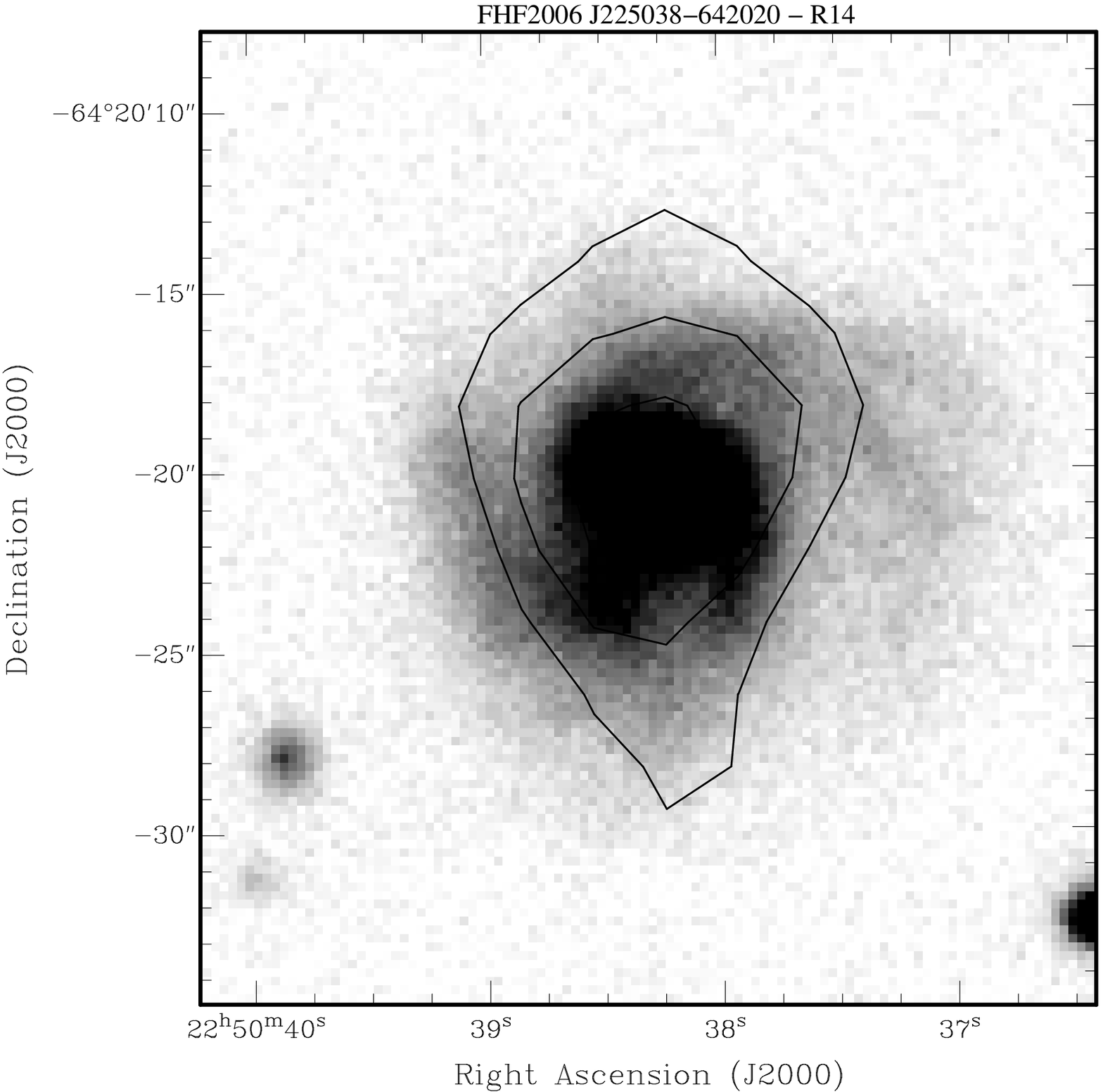}}
\resizebox{12cm}{!}{\includegraphics{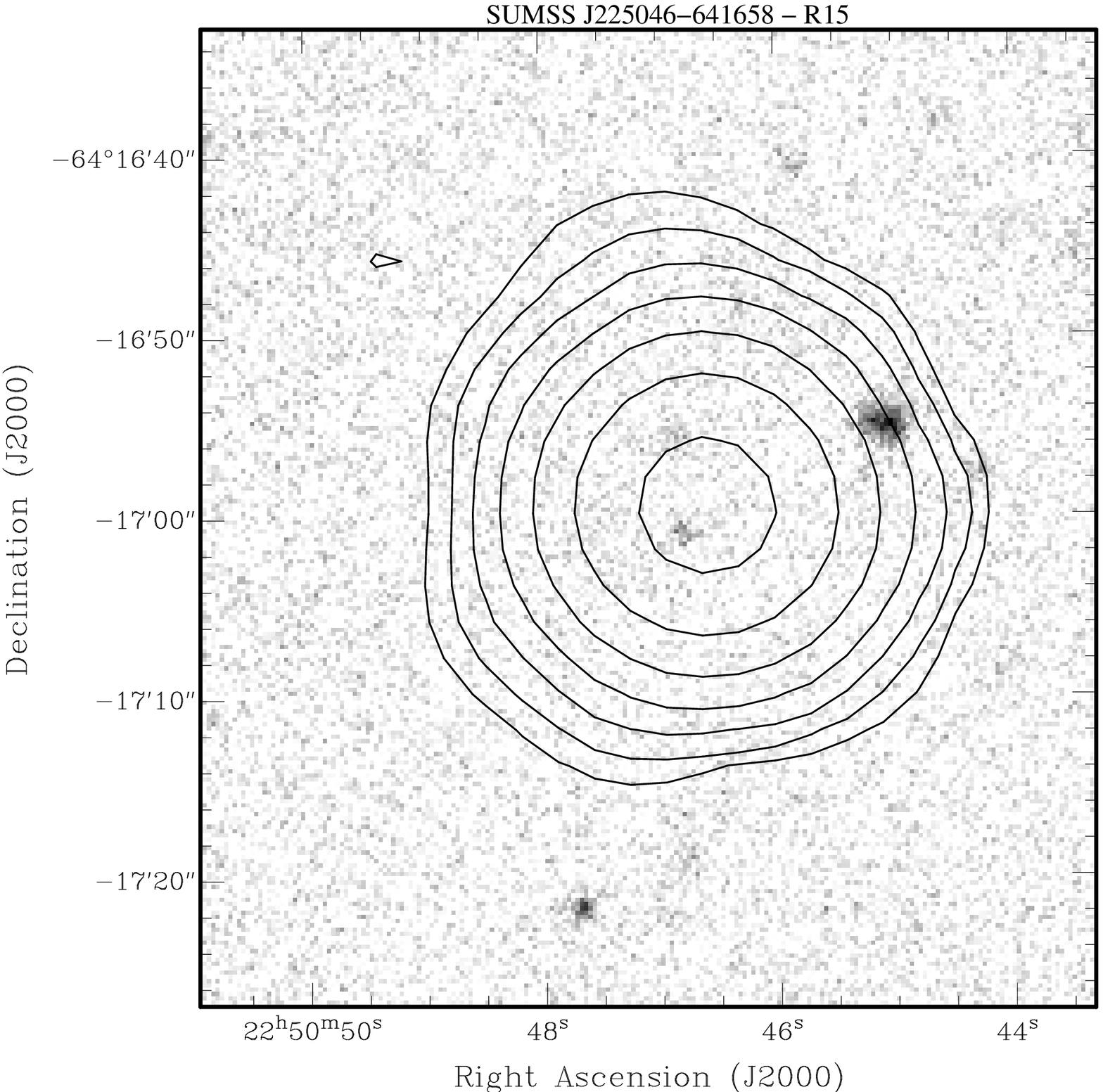}
\includegraphics{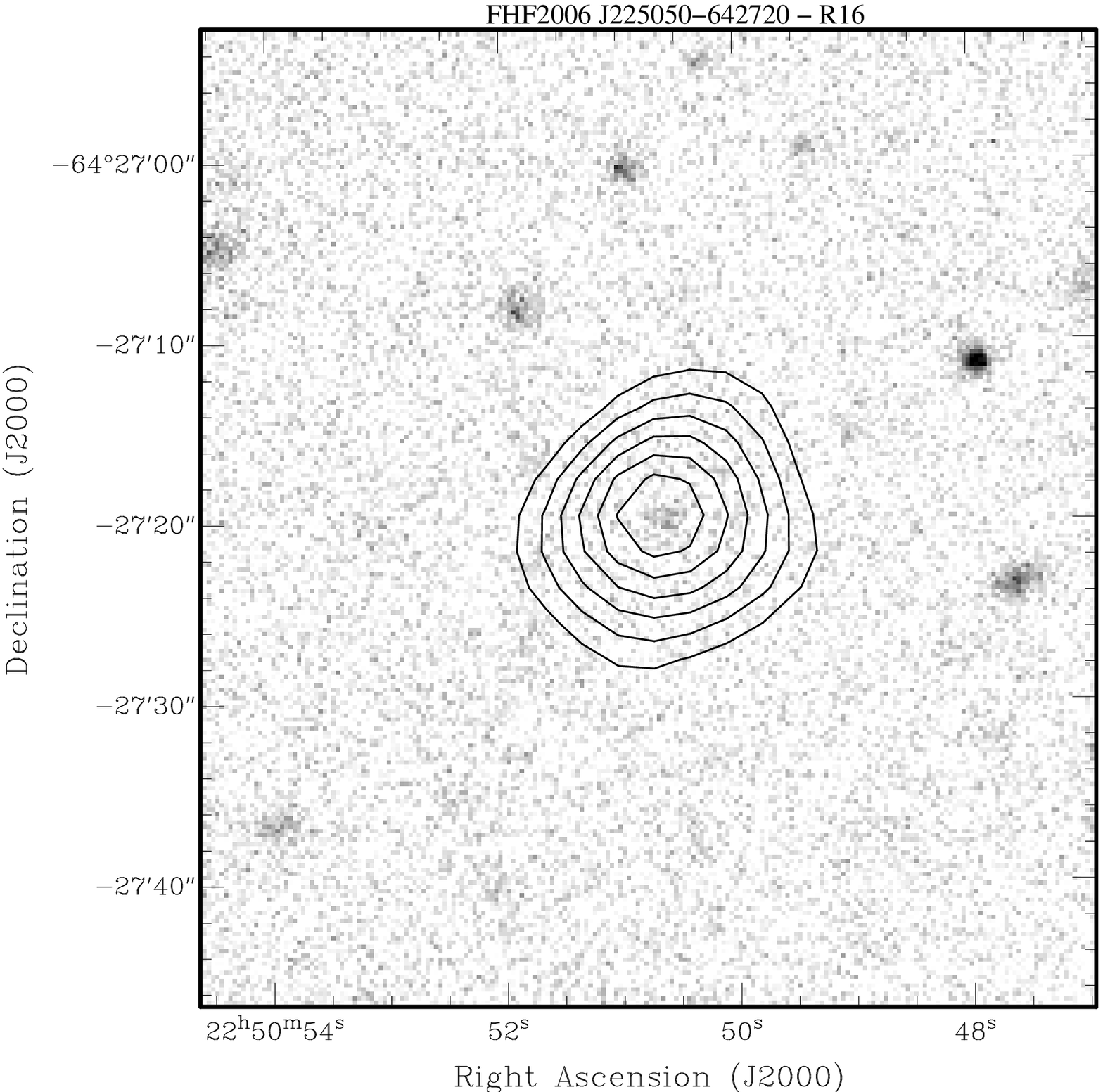}}
\hfill
\begin{center}
\parbox[b]{18cm}{
\caption{Continued. For R12 (FHF2006 J225035$-$642542) radio 
contours are: 0.2205 mJy/beam $\times$ ($-$1, 1, 1.1) (i.e. $\pm$4.5 
and 5$\sigma$ level); for R13 (FHF2006 J225036$-$641608) and R14 
(FHF2006 J225038$-$642020): 0.245 mJy/beam $\times$ ($-$1, 1, 1.2, 
1.4); for R16 (FHF2006 J225050$-$642720): 0.245 mJy/beam $\times$ 
($-$1, 1, 1.4, 1.8, 2.2, 2.6, 3).}
\label{fig:ID}}
\end{center}
\end{figure*}

\noindent The 14 identified radio sources are shown in 
Fig.~\ref{fig:ID} with 22 cm radio contours overlaid on the R-band WFI
images. We considered as reliable identifications all those with
$\mathcal{R}{\leq}3$. For the extended source R9, which shows a
radio-optical offset along the extension of the radio emission and has
$\mathcal{R}>3$ (=12.40), we relied on direct inspection by eye rather
than the value of $\mathcal{R}$. We concluded that R9 is associated
with the bright galaxy (R$\sim$15) on the western side of the radio
emission (see Fig.~\ref{fig:ID}), and that it has a head-tail radio
morphology along the West/East direction. A weak steepening of the
spectrum is observed on the spectral index map of the source moving
from the head towards East. R9 is a confirmed cluster member, located
in the central, high-density region of A3921. Its radio morphology is
therefore probably due to the ram-pressure\footnote{${\rm P}_{\rm
ram}{\propto}\rho_{\rm ICM}*{{\rm v}_{\rm rel}}^2$, where $\rho_{\rm
ICM}$ is the density of the intra-cluster medium, and ${\rm v}_{\rm
rel}$ is the relative velocity between the ICM and the galaxy.}
exerted on the radio emitting plasma by the dense ICM, through which
the galaxy is moving towards the cluster centre (in the East/West
direction).  The extremely peculiar morphology of the radio source R5,
associated with the second BCG, will be discussed in detail in
Sect.~\ref{J2249-6420}.

We also noticed that two sources (R5 and R10) have steep spectral
indexes ($\leq -1.50$). R5 is a confirmed cluster member, while we do
not have redshift information for R10. Based on its magnitude and
morphology, R10 is probably a cluster galaxy: it occupies the region
of A3921 red-sequence on the colour magnitude diagrams ($(V-R)_{\rm
AB}~vs.~R_{\rm AB}$ and $(R-I)_{\rm AB}~vs.~I_{\rm AB}$, F05) and its
morphology and size are very similar to its projected neighbour (see
Fig.~\ref{fig:ID}) which is at the cluster redshift (Katgert et
al. 1996, 1998). The two radio sources with steepest spectra are
therefore very likely cluster members.  Their steep radio spectra may
be related to the pressure of the ICM, which prevents adiabatic
expansion (e.g. Murgia et al. 2004).

\begin{table*}
\begin{center}   
\begin{tabular}{llccccccccc}  
\hline
\hline
\multicolumn{1}{c} {ID$_{\rm R}$} & {ID$_{\rm O}$} & {${\rm RA}_{\rm J2000}$} & {${\rm Dec}_{\rm J2000}$} & {$\Delta_{r-o}$}
& {$\mathcal{R}$} & {V} & {R} & {I} & {$z$} & {${\rm Log} L_{22cm}$} \\
 & &  $^{h~m~s}$ & $ ^{\circ~'~''}$ $ ^{''}$ & $ ^{''}$ &  & mag &  mag & mag &  & (${h_{75}}^{-2}~{\rm W}{\rm Hz}^{-1}$) \\
\hline
R1 & --- & 22 48 35.32 & $-$64 20 38.2 & 0.98 & 1.43 & 22.48 & 22.04 
& 
--- & --- & --- \\
R2 & 57 & 22 48 39.75 & $-$64 19 22.2 & 0.46 & 0.21 & 17.01 & 16.44 
& 15.79 & 0.091 & 21.82 \\
R3 & 69 & 22 48 49.01 & $-$64 23 11.2 & 1.75 & 1.65 & 16.22 & 15.59 
& 14.84 & 0.097 & 22.59 \\
R4 & 215 & 22 49 02.64 & $-$64 27 43.6 & 0.41 & 0.51 & 19.60 & 19.45 
& 19.41 & 0.065$^*$ & 21.56 \\
R5 & 52 & 22 49 04.78 & $-$64 20 35.6 & 1.17 & 1.71 & 15.42 & 14.72 
& 14.00 & 0.093 & 24.00 \\ 
R6 & ---$^{**}$ & 22 49 04.93 & $-$64 18 16.2 & 0.37 & 0.45 & 17.21 
& 16.63 & 15.93 & 0.136$^*$ & 22.93 \\
R9 & 39 & 22 50 06.44 & $-$64 24 42.4 & 8.43 & 12.40 & 16.38 & 15.73 
& 14.99 & 0.094 & 24.12 \\
R10 & --- & 22 50 13.59 & $-$64 19 36.1 & 0.84 & 0.69 & 17.64 & 
17.18 & 16.65 & --- & --- \\
R11 & --- & 22 50 26.50 & $-$64 28 20.1 & 0.69 & 0.86 & 17.69 & 
17.24 & 16.67 & --- & --- \\
R12 & --- & 22 50 35.40 & $-$64 25 46.8 & 4.24 & 1.64 & 20.56 & 
19.21 & 18.22 & --- & --- \\ 		 
R13 & --- & 22 50 36.62 & $-$64 16 07.6 & 1.05 & 0.58 & --- & 20.57 
& 19.56 & --- & --- \\
R14 & ---$^{**}$ & 22 50 38.22 & $-$64 20 21.1 & 1.00 & 0.49 & 16.73 
& 16.33 & 15.88 & 0.093$^*$ & 22.20 \\
R15 & --- & 22 50 46.78 & $-$64 17 00.9 & 1.88 & 2.77 & --- & 23.22 
& --- & --- & --- \\
R16 & --- & 22 50 50.62 & $-$64 27 19.8 & 0.29 & 0.31 & --- &  22.92 
& 
21.35 & --- & --- \\
\hline
\end{tabular}
\caption{Optical identification of the $>5\sigma$ radio sources. 
``---'' indicates unknown values. R5 and R3 correspond to BG2 and BG3
respectively in F05. $^*$ indicates uncertain $z$ (Q.F.$>1$ in
F05). $^{**}$ z from ENACS catalogue, not from F05.}
\label{tab:ID}
\end{center}
\end{table*}

\section{Star-forming galaxies and AGNs in the A3921 central field}\label{SF-AGN}

Of the 14 radio sources with optical identification, spectroscopic
redshifts are available for seven of the host galaxies; of these, five
lie at the cluster redshift. Due to their colour and size, we
considered the two additional radio identifications without redshift
information (R10 (see Sect.~\ref{ROid}) and R11) as likely cluster
member candidates. This section concentrates upon these 7 radio
emitting cluster members (confirmed or candidate) and upon the
star-forming and post-star-forming galaxies identified in the optical
(F05).

The optical spectroscopic analysis of the A3921 central field 
revealed the presence of 24 galaxies with spectral signatures of 
recent or ongoing SF (F05). The spectral classification was based on 
the presence and strength of the ${\rm H}_\delta\lambda$4101 
and [OII]$\lambda$3727 lines, observed in absorption and emission, 
respectively. Galaxies presenting an intermediate and strong ${\rm 
H}_\delta$ but no [OII] emission were classified as 
post-star-burst/post-star-forming (k+a, 13 objects). The presence of 
[OII] emission was considered as a possible hint for ongoing SF. F05 
found 11 emission line galaxies that, based on the MORPHS 
classification (Dressler et al. 1999; Poggianti et al. 1999), were 
divided into: i) two e(b)-type objects (i.e. possible star-burst 
galaxies, Poggianti et al. 1999), with strong [OII] emission 
(EW$<-5$\AA), ii) four e(c)-type galaxies, with weak Balmer 
absorption and weak to moderate [OII] emission, classified as 
spirals forming stars at a constant rate by Poggianti et al. (1999), 
and iii) five e(a)-type objects, with strong ${\rm H}_\delta$ 
absorption and measurable [OII] emission, suggested to be either 
dusty star-burst or post-star-burst galaxies with some residual SF 
(Poggianti et al. 1999). They are given in Table~\ref{tab:SFR} 
(upper part). Most of the emission line galaxies are concentrated in 
the region of A3921-B and in between the two main sub-clusters A and 
B (see Fig. 22 in F05 and top panel of 
Fig.~\ref{fig:FOC_radio_ottico} in the present paper), suggesting 
that the ongoing merging event in A3921 could have triggered their 
SF/star-burst episode.

However, the [OII] emission line as an estimator of SF is fraught 
with problems (e.g. Kennicutt 1992), the most important of which is 
that it suffers from strong dust extinction. As a consequence, the 
SF activity can be strongly underestimated (or even not detected) if 
only determined with [OII]. Low-power radio emission from galaxies 
is, on the contrary, an extinction-free tracer of SF. Moreover, in 
our case the spectral range covered by the optical observations 
($\sim$ 3000--6000 \AA) was not sufficient to discriminate between 
star-forming/bursting objects and AGNs among the emission-line 
galaxies. In order to test if the merging event has affected the 
star formation history (SFH) of A3921 galaxies we still need to:

\begin{itemize}

\item divide emission line galaxies into star-forming/bursting objects 
and AGN;

\item verify that the emission line galaxies are really 
located in the collision region and not infalling field 
galaxies seen in projection at the cluster centre. In such a case 
the merging event would not be affecting either the SF or the 
nuclear activity of these objects;

\item test the hypothesis that all the detected k+a-type objects are 
in a {\bf post}-star-bursting/{\bf post}-star-forming phase. k+a 
galaxies could also be strongly obscured star-burst galaxies (e.g. 
Smail et al. 1999; Poggianti \& Wu 2000), in which the [OII] line is 
totally extinguished. In such a case, star-bursting galaxies would 
not be concentrated only in the collision region of A3921, but 
spread all over the cluster (see Fig.~18 of F05). The connection 
between the sub-clusters' merger and the on-going star-burst episode 
would obviously be less probable.

\end{itemize} 

\noindent We thus combined the optical and radio observations in order to
measure the SFR of the cluster members independently from dust
extinction and classify the active galaxies of the A3921 central field
either as star-forming/bursting objects or AGNs.

\subsection{Emission-line galaxies: optical/radio comparison}

\subsubsection{Optical and radio SFR}

To obtain star-formation rates from radio luminosities, we used the
relation from Condon (1992), which was estimated assuming solar
abundances and a Salpeter initial mass function (IMF) above 5
M$_{\odot}$. In order to estimate the SFR of stars more massive than
0.1 M$_{\odot}$, we divided Condon's relation by 0.18 (e.g. Serjeant
et al. 2002):

\begin{equation}
\left[\frac{{\rm SFR}(M{\geq}0.1{\rm M}_{\odot})}{{\rm 
M}_{\odot}{\rm 
yr}^{-1}}\right]~=~1.1{\times}10^{-21}~\left(\frac{L_{\nu}}{\rm W/Hz}\right)~\left(\frac{\nu}{\rm GHz}\right)^{-\alpha}
\end{equation}

\noindent where $L_{\nu}$ is the non-thermal luminosity at the frequency $\nu$
(in our case 1.344~GHz), and ${\alpha}$ is the non-thermal spectral
index. The SFR of the 11 emission line galaxies detected in A3921
central field are summarised in Table~\ref{tab:SFR}.

We also estimated the SFR of the star-forming galaxies in A3921 from 
their [OII] emission, adopting the relation by Kennicutt 
(1998)\footnote{SFRs from EW([OII]) had already been estimated in 
F05. Here we re-estimated these SFRs with the same IMF adopted in 
the previous radio relation.}:

\begin{eqnarray}
\left[\frac{{\rm SFR}(M{\geq}0.1{\rm M}_{\odot})}{{\rm 
M}_{\odot}{\rm yr}^{-1}}\right] & = & 
1.4{\times}10^{-34}~\left(\frac{L_{\rm [OII]}}{\rm W}\right) \nonumber \\
 & \simeq & 2.0{\times}10^{-12}~\frac{L_B}{L_{B_{\odot}}}~EW(\rm [OII])E({\rm H}_{\alpha})
\label{eqn:SFR-OII}
\end{eqnarray}

\noindent where we used the relation by Kennicutt (1992) to 
determine the luminosity of the [OII] emission line from a) its
equivalent widths (EW(\rm [OII])), b) the integrated broad-band $B$
luminosity of the galaxy in solar units ($L_B/L_{B_{\odot}}$), and c)
a reasonable value for the extinction correction ($E({\rm
H}_{\alpha})$). In Table~\ref{tab:SFR} we give the SFR from EW([OII])
both with and without a correction for dust-extinction. When
included, a canonical value for extinction (1 mag for ${\rm
H}_{\alpha}$) has been applied. In Eq.~\ref{eqn:SFR-OII} the errors
can be considerably large ($\pm$~30--50\%), mostly due to the adopted
relation between the B-band and [OII] line luminosities (Kennicutt
1992).

The results are summarised in the upper part of Table~\ref{tab:SFR},
which lists:

\begin{itemize}

\item[-] {\it Columns (1) \& (2):} optical and radio ID (corresponding 
to Table~A in F05 and Table~\ref{tab:5s} in this
paper, respectively).

\item[-] {\it Columns (3) \& (4):} optical coordinates.

\item[-] {\it Columns (5) \& (6):} redshift and spectral type of the source.

\item[-] {\it Column (7):} Log of the absolute spectral 
luminosity at 1.344~GHz (W/Hz). For the sources not detected at radio
wavelengths, we estimated an upper limit (at 3$\sigma$ significance
level) using $S_{1.344~GHz}<$0.147 mJy/beam and $\alpha$=$-$0.8.

\item[-] {\it Columns (8) \& (9):} SFR of the source estimated from its 
[OII] line luminosities, without ($E({\rm H}_{\alpha})=0$) and with
($E({\rm H}_{\alpha})=1$) the extinction correction.

\item[-] {\it Column (10):} SFR estimated from the 1.344~GHz luminosity of 
the source.

\end{itemize}

\begin{table*}
\begin{center}   
\begin{tabular}{llcccccccc}  
\hline
\hline
${\rm ID}_{\rm O}$ & ${\rm ID}_{\rm R}$ & ${\rm RA}$ & ${\rm Dec}$ &
 $z$ & Sp. Type & ${\rm Log} L_{22cm}$ & ${{\rm
 SFR}_{[OII]}}^{\diamondsuit}$ & ${{\rm
 SFR}_{[OII]}}^{{\diamondsuit}{\diamondsuit}}$ & ${\rm SFR}_{\rm
 1.4~GHz}$\\
& & (J2000) & (J2000) & & & (${h_{75}}^{-2}{\times}$ &
 (${h_{75}}^{-2}{\times}$ & (${h_{75}}^{-2}{\times}$ &
 (${h_{75}}^{-2}{\times}$ \\
\multicolumn{1}{c}{\#} & {\#} & {$^{h~m~s}$} & {$ ^{\circ~'~''}$} &{ 
} &{ }& {${\rm W}/{\rm Hz}$}) & {M$_{\odot}{\rm yr}^{-1}$)} &  
{M$_{\odot}{\rm yr}^{-1}$)} & {M$_{\odot}{\rm yr}^{-1}$)} \\ \hline
17 & --- & 22 48 59.0 & $-$64 21 11.8 & 0.094 & e(a) & $<$21.42 & 
0.28 & 0.70 & $<$3.68 \\
45 & --- & 22 49 41.5 & $-$64 26 24.1 & 0.085 & e(c) & $<$21.33 & 
0.04 & 0.11 & $<$2.98 \\
57 & R2 & 22 48 39.8 & $-$64 19 22.2 & 0.091 & e(c) & 21.82 & 0.25 & 
0.62 & 7.94 \\
69 & R3 & 22 48 49.0 & $-$64 23 11.2 & 0.097 & e(b) & 22.59 & 5.64 & 
14.09 & 50.35 \\
73 & --- & 22 48 34.8 & $-$64 23 39.9 & 0.095 & e(c) & $<$21.43 & 
0.26 
& 0.64 & $<$3.76 \\
81 & --- & 22 49 41.1 & $-$64 24 05.6 & 0.097 & e(a) & $<$21.45 & 
0.05 
& 0.12 & $<$3.93 \\
82 & --- & 22 49 38.4 & $-$64 23 23.8 & 0.099 & e(a) & $<$21.47 & 
0.16 
& 0.40 & $<$4.11 \\
100 & --- & 22 49 41.6 & $-$64 19 59.0 & 0.086 & e(b) & $<$21.34 & 
0.70 & 1.74 & $<$3.05 \\
169 & --- & 22 50 01.8 & $-$64 22 21.8 & 0.085 & e(a) & $<$21.33 & 
0.12 & 0.31 & $<$2.98 \\
181 & --- & 22 49 26.1 & $-$64 23 21.4 & 0.095 & e(c) & $<$21.43 & 
0.10 & 0.24 & $<$3.76 \\
226 & --- & 22 49 12.0 & $-$64 16 03.9 & 0.100 & e(a) & $<$21.48 & 
0.17 & 0.43 & $<$4.20 \\
\hline
\hline
39 & R9 & 22 50 06.4 & $-$64 24 42.4 & 0.094 & k & 24.12 &  &  & \\
52 & R5 & 22 49 04.8 & $-$64 20 35.6 & 0.093 & k & 24.00 & &  \\ 
---$^{*}$ & R14 & 22 50 38.2 & $-$64 20 21.1 & 0.093 & --- & 22.20 & & \\
--- & R10 & 22 50 13.6 & $-$64 19 36.1 & --- & --- & 22.26 &  &  & 
\\
--- & R11 & 22 50 26.5 & $-$64 28 20.1 & --- & --- & 22.67 &  &  & 
\\
\hline
\end{tabular}
\caption{Cluster members with emission lines (and possible radio 
emission) ({\bf top}), and with radio emission but no emission lines
({\bf bottom}). The optical ID corresponds to Table A in F05. The
coordinates are those of the optical position. {\bf Top:} The SFR of
the emission line cluster members was measured through their [OII]
($^{\diamondsuit}$: extinction correction term $E({\rm
H}_{\alpha})=0$.  $^{{\diamondsuit}{\diamondsuit}}$: extinction
correction term $E({\rm H}_{\alpha})=1$.) and 1.4~GHz luminosities.
In case of no radio detection an upper limit (3$\sigma$) was
estimated. {\bf Bottom:} Confirmed and candidate cluster members
detected at radio wavelengths and either without optical emission
lines or unknown spectral type. Among the objects with available $z$,
spectral types are available neither for poor S/N (Q.F.$>$1 in F05)
nor for ENACS (Katgert et al. 1996 and 1998; Mazure et al. 1996)
spectra. For candidate cluster members we assumed the mean cluster
redshift. $^{*}$: object from the ENACS catalogue.  }
\label{tab:SFR}
\end{center}
\end{table*}

\noindent These results are discussed in the following section.

\subsubsection{Emission line galaxies with detected radio emission}\label{EL-radio}

Among the 11 emission-line cluster members detected in A3921 by F05,
only two have detectable radio emission at our flux limit (0.15
mJy/beam, 3$\sigma$ level).

One (R2) is very likely a classical spiral galaxy, as suggested by 
most of its observed properties. Firstly, based on the presence and 
strength of [OII] emission and ${\rm H}_{\delta}$ absorption lines, 
F05 classified R2 as an e(c)-type object (MORPHS classification; 
Dressler et al. 1999). Following Poggianti et al. (1999), e(c) 
spectra are typical for present-day spirals. This conclusion agrees 
with the optical morphology and the radio SFR of R2, typical of 
late-type objects. The red colour of this galaxy and the low 
SFR$_{\rm [OII]}$ (significantly lower than the radio estimate) 
could be due to the fact that, in the optical, the younger 
(disk) component of the galaxy is much more obscured by dust lanes 
than the old bulge.

The other emission-line galaxy detected at radio wavelengths (R3) is
associated with the third brightest cluster galaxy (BG3 in F05), and
it is located at the barycentre of the X-ray emission of A3921-B
(B05). In optical it is characterised by: a) an elliptical morphology;
b) red $({\rm V-R})_{AB}$ and $({\rm R-I})_{AB}$ colours; c) a high
radial velocity offset from the mean cluster redshift (+788 km/s); d)
strong [OII] emission ({\it EW}([OII])=$-$65 \AA). Based on its
spectral properties BG3 was classified as a star-burst galaxy by F05,
but, based on its nebular emission line ([OII]), it could also be a
Seyfert or a LINER.  Unfortunately, the very limited spectral range
covered by the optical observations did not allow F05 to use any
diagnostic diagrams to distinguish between these two hypotheses.

\begin{figure}
\begin{center}
\resizebox{9cm}{!}  
{\includegraphics{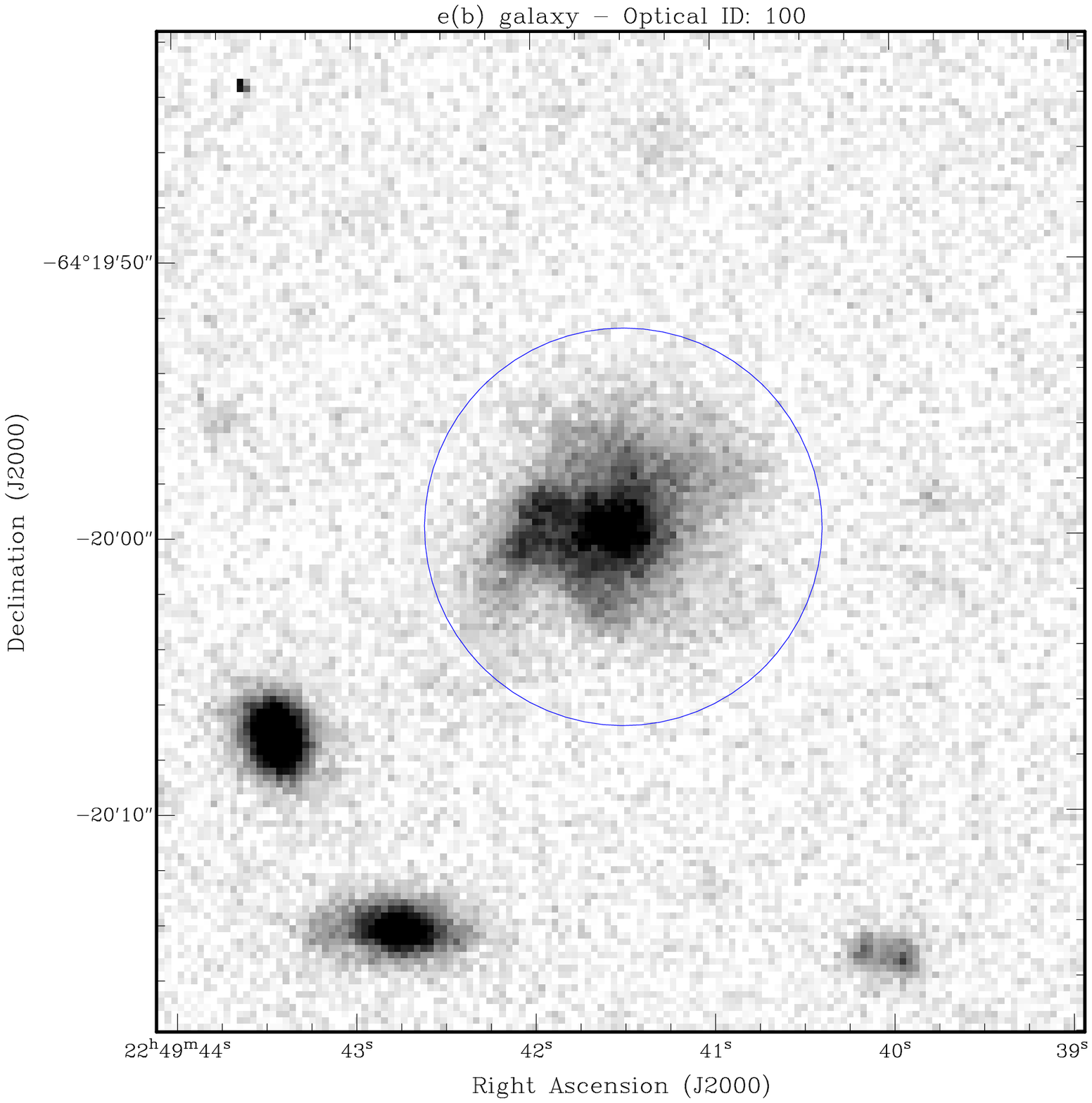}}
 \hfill  
\parbox[b]{9cm}{
\caption{R-band image of the e(b) galaxy with optical ID=100 in 
Table~\ref{tab:SFR}.}
\label{fig:eb_100}}
\end{center}
\end{figure}

We therefore analysed in detail the radio and X-ray (B05) properties
of BG3. Following Morrison \& Owen (2003) and correcting for our
cosmology, we considered as AGNs the radio galaxies with
Log~($L_{1.4~GHz}{\rm (W/Hz)})\gtrsim$22.73, and as star-forming
objects those below this limit. The radio luminosity of BG3 (${\rm
Log}~(L_{22 cm}{\rm (W/Hz)})$=22.59) is on the upper side of the
luminosity distribution of star-forming galaxies, albeit very close to
the boundary with AGNs. Thus, from a statistical point of view, BG3
has radio properties typical of star-forming/bursting objects, but we
cannot completely exclude the faint AGN hypothesis.

The XMM analysis of A3921 by B05 revealed a diffuse extended X-ray
emission for BG3, with a thermal spectrum.  This points against an AGN
origin of the emission, and we suggest that the intense star-burst
episode observed in BG3 has probably injected hot gas into the
galaxy's inter-stellar medium (ISM). Both the red colours and the
strong difference between the optical and radio SFR estimates (see
Table~\ref{tab:SFR}) could be due to dust extinction.  Actually, it
has been shown that, in clusters, star-forming galaxies heavily
affected by dust extinction are more centrally concentrated than
star-forming galaxies in general (Miller \& Owen 2002). MIR
observations also revealed a higher fraction of dusty-star-burst
galaxies in dynamically young clusters (Coia et al. 2005), as in the
case of A3921. Therefore, we concluded that the red colours of R2 and
R3 are probably due to their dusty star-forming/bursting nature
(e.g. Wolf, Grey \& Meisenheimer 2005).

\subsubsection{Emission line galaxies without detected radio emission}

None of the other nine emission line cluster members were 
detected at the 3$\sigma$ sensitivity limit of our radio 
observations. These galaxies are thus characterised by SFRs lower 
than $3-4~{{\rm M}_{\odot}{\rm yr}^{-1}}$.

We suggest that the SF properties of the emission-line galaxies 
detected in the collision region of A3921 may be due to cluster 
environmental effects. Most of them are brighter than the 
``dwarfs/giants'' magnitude cut applied by Poggianti et al. (2004) 
in their analysis of the Coma cluster (see F05), and they all have 
SFRs lower than those typical of gas-rich spirals ($\sim20~{{\rm 
M}_{\odot}{\rm yr}^{-1}}$, e.g. Kennicutt 1998). They could be 
gas-poor objects, whose gas deficiency is related to cluster 
environmental effects, and in particular to hydrodynamical 
interactions between the galaxy ISM and the ICM (ram-pressure and 
viscous stripping, thermal evaporation; see Boselli \& Gavazzi 2006 
for a recent and complete review).

The sub-cluster merger in A3921 could have subsequently enhanced the 
SF activity of the galaxies located in the collision region. It has 
been shown that both dynamical interactions with the ICM and tidal 
interactions with nearby companions and/or the cluster potential can 
induce SF in cluster galaxies (e.g. Gavazzi et al. 1995; Moss \& 
Whittle 2000; Bekki 1999). Actually, these physical mechanisms are 
particularly effective in merging clusters since: a) the ISM 
compression due to galaxy-ICM interactions is most efficient where 
the density and the temperature of the ICM reach their maximum (i.e. 
around the ICM compression bar in A3921 detected by B05), and b) the 
time-varying gravitational potential of merging clusters enhances 
the rate of galaxy encounters and drives efficient transfer of gas 
to the central region of galaxies, triggering a starburst.

We can extract more physical information about some of the 
emission-line galaxies without radio emission at the 3$\sigma$ level 
by comparing in more detail their optical and radio properties. One 
of these objects has spectral features typical of star-bursting 
galaxies (ID$_{\rm O}$=100 in Table~\ref{tab:SFR}), i.e. strong 
[OII] emission (EW$<-$40 \AA). Its distorted optical morphology 
(Fig.~\ref{fig:eb_100}) suggests that the observed star-burst 
episode could be due to interaction with another galaxy.

Three galaxies have spectra typical of spirals forming stars at a
constant rate (e(c)), while the remaining 5 belong to the e(a) class,
that, based on spectral modelling (Poggianti et al. 1999), could be
either dusty star-burst or post-star-burst galaxies with some residual
SF. Up to now only e(a) galaxies belonging to the first class have
been detected, showing extremely strong dust extinction factors
(median value $\sim$12 mag, Duc et al. 2002). For our e(a) objects, by
dividing the 8$^{th}$ and 10$^{th}$ columns of Table~\ref{tab:SFR}, we
find SFR$_{\rm 1.4GHz}$/SFR$_{[OII]}<$13--79, corresponding to dust
extinction values $<$5--30 mag (lower than 12 mag for 4/5 of the
sample).  This means that, for extinction higher than these lower
limits, we should observe the e(a) galaxies if they were dusty
star-burst. Since the estimated extinction factors are comparable to
those found by Duc et al. (2002) and we detected none of the e(a)
galaxies at radio wavelengths, we suggest that at least a fraction of
them could not be dusty star-burst, but post-star-burst galaxies with
some residual SF. This hypothesis has to be tested with more sensitive
observations (e.g. higher-resolution spectroscopy covering a wider
wavelength range, IR and deeper radio observations) as our SFR
determination has big errors. The highest uncertainty is related to
the SFR derived from the [OII] line, which suffers from several
problems other than extinction (e.g. slit aperture and metallicity
effects).

\subsubsection{k+a galaxies}

Among the cluster members observed by F05, 13 are characterised by an
intermediate or strong ${\rm H}_{\delta}$ line in absorption and no
detectable [OII]3727 line in emission (k+a galaxies; Dressler et
al. 1999). These galaxies are generally interpreted as objects that
have no significant ongoing SF, but that were forming stars in the
recent past ($\lesssim$1-1.5 Gyr). They are therefore usually
classified as post-star-forming (moderate Balmer lines, EW(${\rm
H}_{\delta})<$4--5 \AA) or post-star-bursting objects (strong Balmer
lines, EW(${\rm H}_{\delta})>$4--5 \AA). However, it has been
suggested that k+a (like e(a)) might also be star-bursting galaxies
whose emission lines are invisible in the optical due to heavy or
selective dust extinction (e.g. Smail et al. 1999; Poggianti \& Wu
2000; Goto 2004). Understanding the SF properties of the k+a galaxies
in A3921 is important firstly to address the debated nature of this
class of objects, and secondly to confirm (or reject) the spatial
correlation between the detected star-forming objects and the
collision region of the cluster.

None of our target k+a galaxies was detected down to our 2$\sigma$
sensitivity limit (SFR$\sim3 {{\rm M}_{\odot}{\rm yr}^{-1}}$ at the
mean redshift of the cluster and assuming $\alpha$=$-$0.8). This means
that no galaxies of this spectral class host dust obscured SF activity
up to our $2\sigma$ sensitivity limit ($\sim3{M_{\odot}{\rm
yr}^{-1}}$). Based also on the moderate ${\rm H}_{\delta}$ equivalent
width and on the red colours of the k+a galaxies detected in A3921
(F05), we can therefore conclude that they are not strong dusty
star-burst, and, in agreement with the results found by Duc et
al. (2002) in the cluster A1689, we suggest that these objects are in
a post-star-forming phase as predicted by their spectral
properties. Due to the limit in sensitivity of our radio observations,
we cannot, however, totally exclude the possibility that they are
dust-obscured galaxies forming stars at a very low rate. The biggest
issue in this respect is that the spectroscopic classification of
these objects is based solely on the [OII] emission and Balmer
absorption lines. The importance of having information as well on the
${\rm H}_{\alpha}$ line has been shown (Goto 2004). An ongoing
spectroscopic follow-up covering a wider spectral range will allow us
to definitively test the post-star-forming nature of the k+a galaxies
in A3921.

\begin{figure}
\begin{center}
\resizebox{9cm}{!}  
{\includegraphics{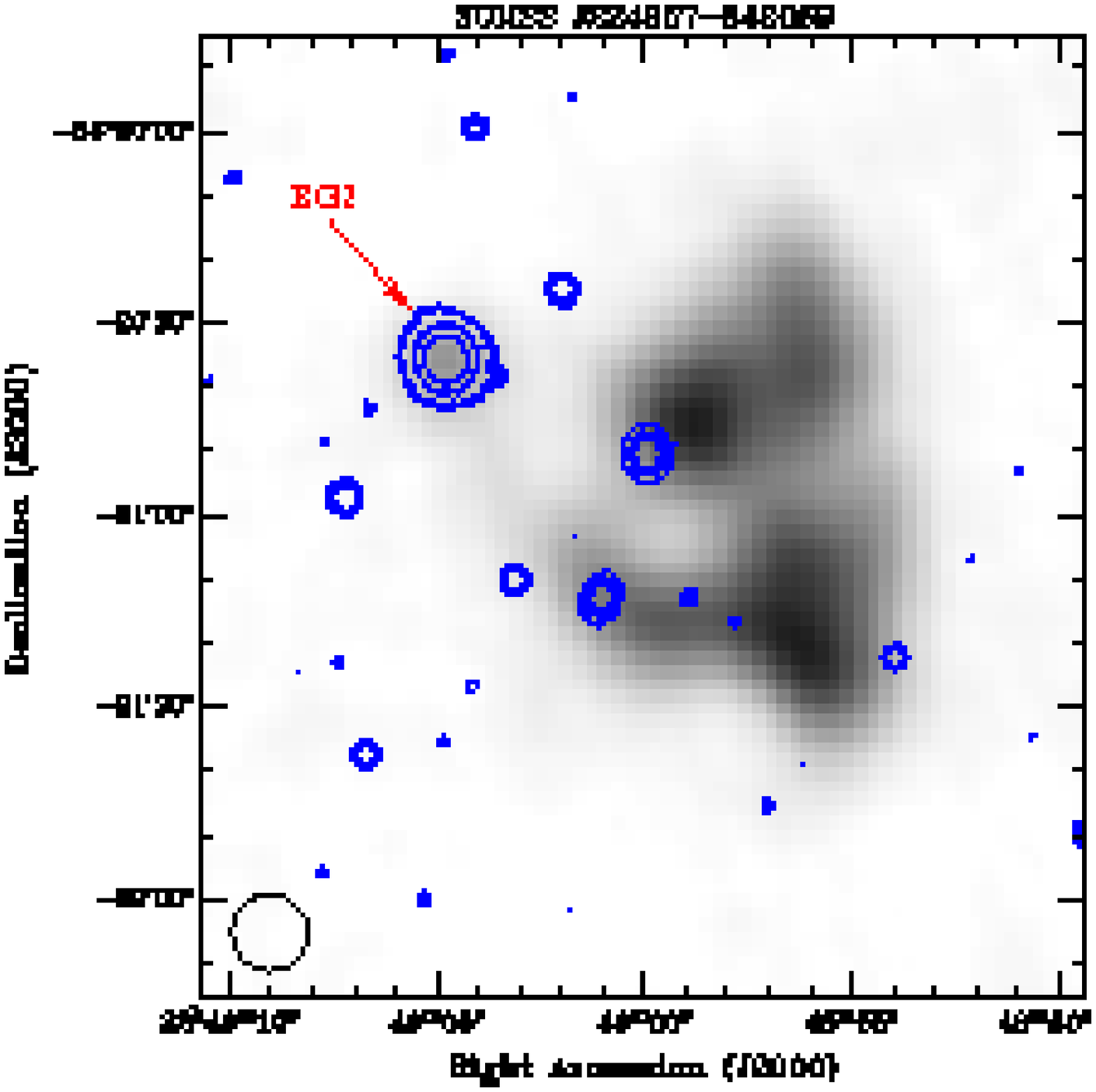}}
 \hfill  
\parbox[b]{9cm}{
\caption{R-band contours overlaid on the 1.344~GHz image of 
the peculiar radio source SUMSS J224857$-$642059 (R5).  The beam 
size (shown in the bottom-left corner) is $12''{\times}12''$.}
\label{fig:grey22}}
\end{center}
\end{figure}

\begin{figure*}
\begin{center}
\resizebox{18cm}{!}{
\includegraphics{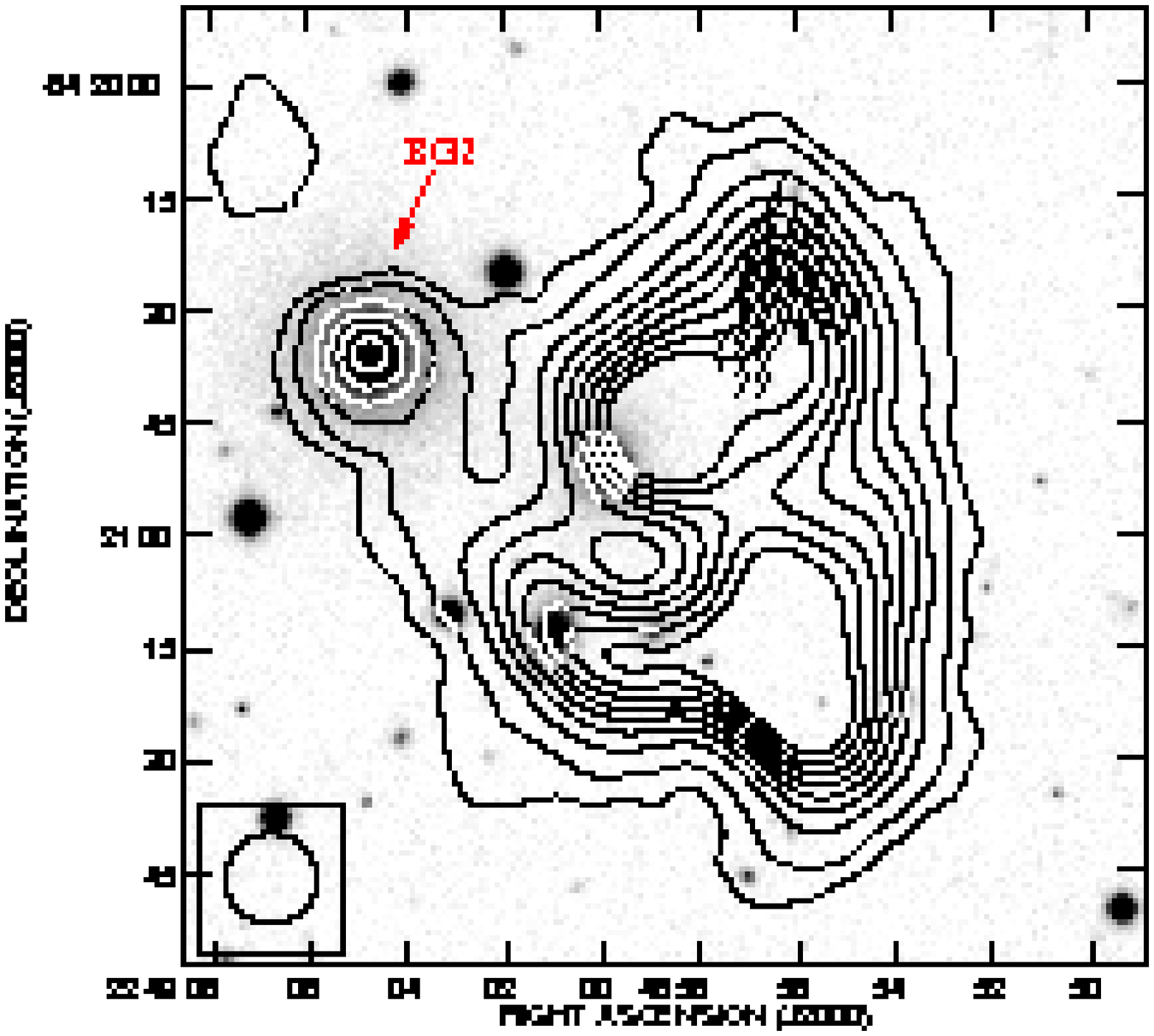}
\includegraphics{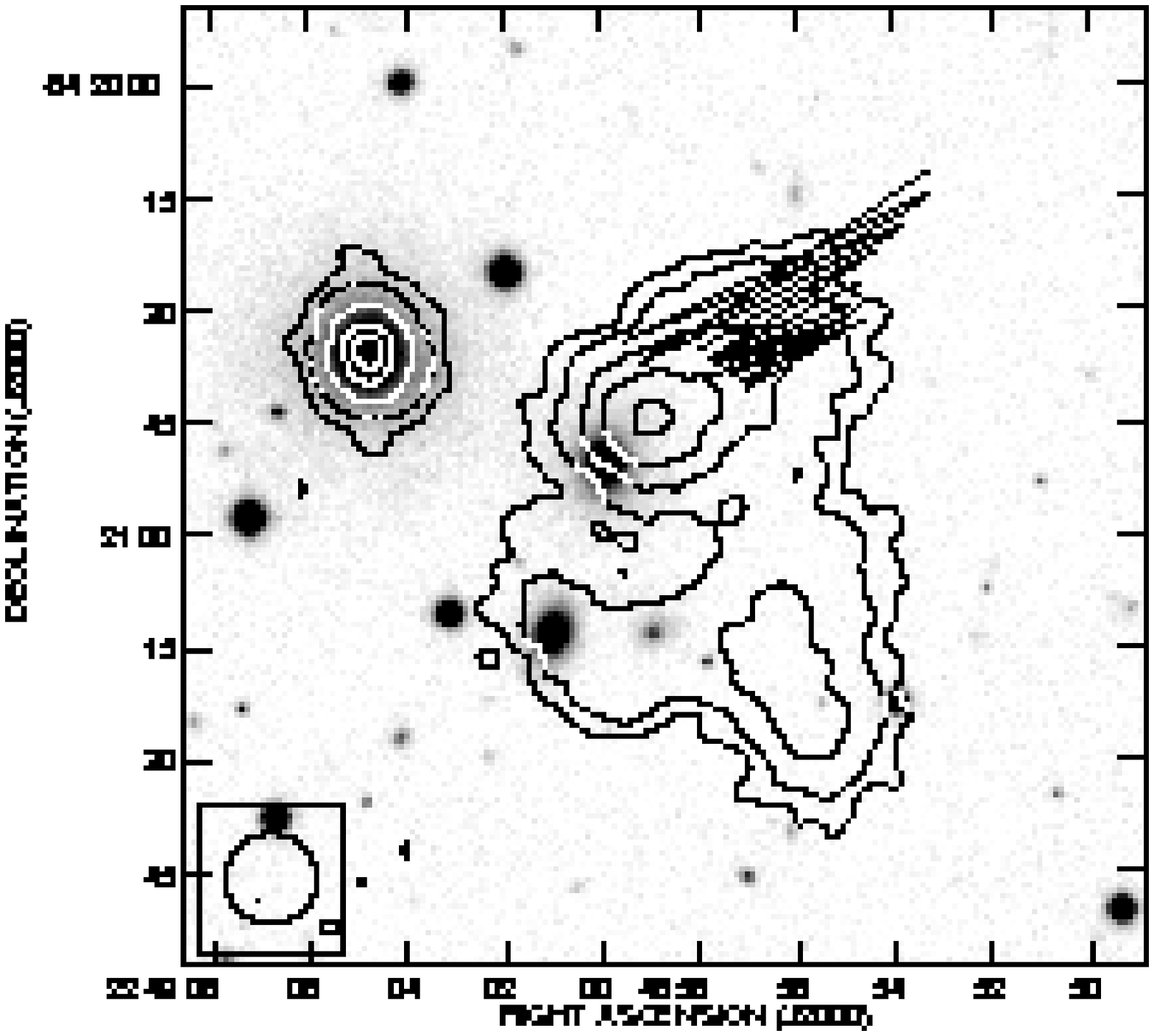}}
 \hfill  
\parbox[b]{18cm}{
\caption{1.344~GHz (left) and 2.368~GHz (right) radio contours of the 
extended radio source SUMSS J224857-642059 (R5), superimposed on the
R-band optical image (WFI observations, F05). The radio emission is
associated with the second brightest cluster galaxy (BG2 in F05). In
both images the angular resolution is $12''{\times}12''$. {\bf Left:}
the contour levels are: 0.147 mJy/beam $\times$ (-1, 1, 2, 4, 6, 8,
10, 12, 14, 16). The rms noise level of the map is 0.049 mJy/beam.
{\bf Right:} the contour levels are: 0.159 mJy/beam $\times$ (-1, 1,
2, 4, 6, 8, 10, 12, 14, 16). The rms noise level of the map is 0.053
mJy/beam. In both panels, superimposed lines represent the orientation
of the electric vector ({\it {\bf E}}) and are proportional in length
to the fractional polarization (1$''$=1.40\%).}
\label{fig:BG2-RS5}}
\end{center}
\end{figure*}

\begin{figure*}
\begin{center}
\resizebox{16cm}{!}  
{\includegraphics{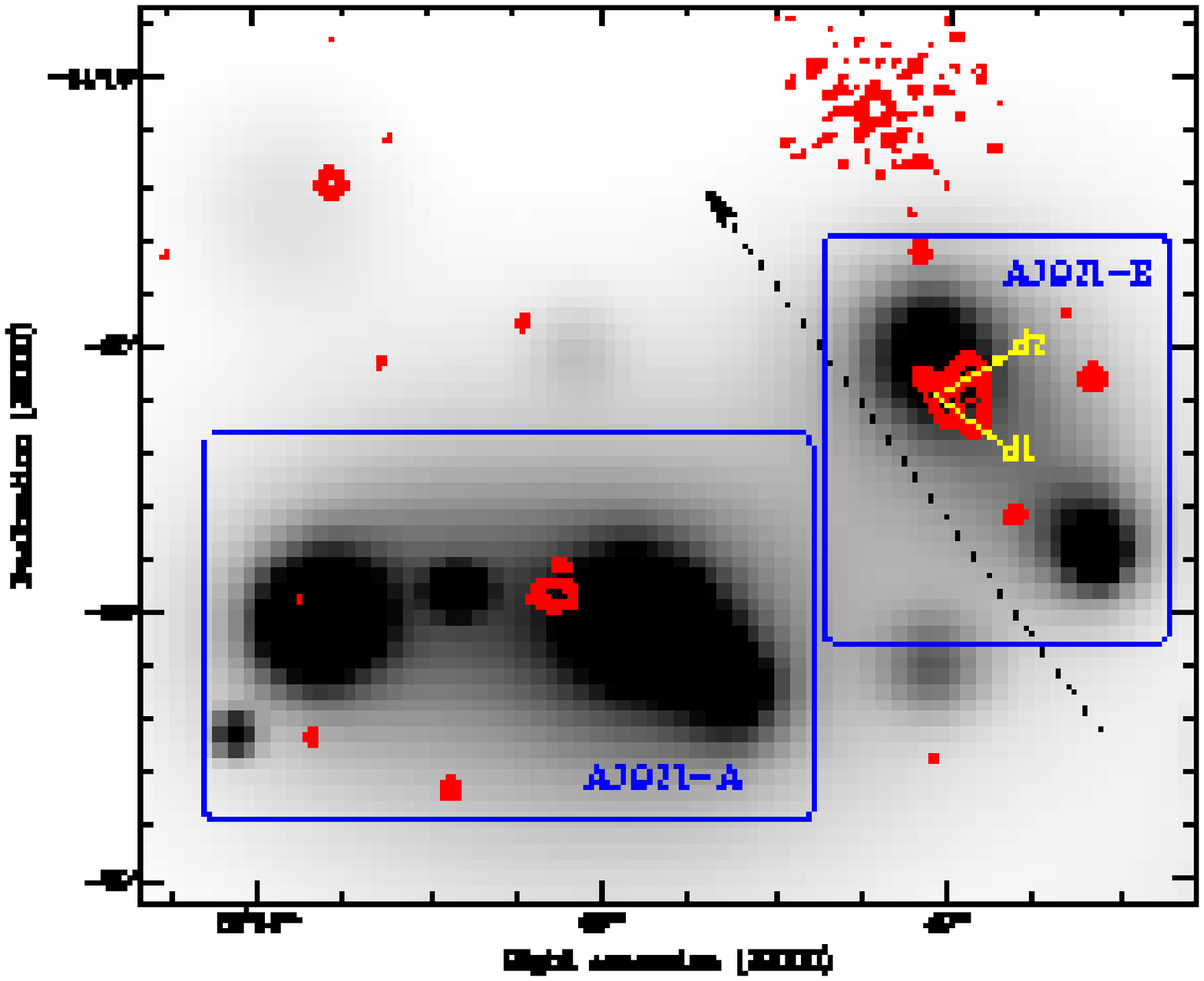}} 
\hfill
\parbox[b]{18cm}{
\caption{1.344~GHz radio contours (levels from 0.294 to 15.680 mJy/beam, 
with an increment factor of 2) overlaid on the projected galaxy
density map of the galaxies lying in A3921 red sequence (F05). The two
main sub-clusters of A3921 (A3921-A and A3921-B) and the main
elongation axis of the radio tail of SUMSS J224857$-$642059 (R5) (d1 
and
d2) are indicated. The off-axis merging direction of A3921-B
reconstructed from the combined optical and X-ray analyses is also
shown with a dashed arrow (F05 and B05).}
\label{fig:clumps}}
\end{center}
\end{figure*}

\subsection{Radio galaxies}

In the bottom panel of Table~\ref{tab:SFR} we list the five confirmed
or candidate cluster members i) with detected radio emission, and ii)
without emission lines, or for which the spectral type is
unknown. Three of them (R10, R11 and R14) are located in the Eastern
region of A3921-A (i.e. not in the collision region), and they are
very likely gas rich, dusty star-forming late-type galaxies, since
they have: i) radio luminosities below the AGN/SF boundary (Log
($L_{1.4~GHz}{\rm (W/Hz)})=$22.73, see Sect.~\ref{EL-radio}), ii)
spiral-like morphologies (R14 also shows a disturbed morphology,
suggesting that the SF activity could be triggered by galaxy-galaxy
interaction), and iii) red colours probably due to dust
extinction. The remaining two radio emitting cluster members (R9 and
R5) are on the contrary AGNs, their optical morphology and spectrum
being typical of early-type galaxies and their radio luminosity being
well above the AGN/SF boundary.  We therefore detected only two AGNs
in the central field of A3921.  A similar number of AGNs was
detected in the nearby clusters Coma and A1367 (Gavazzi \& Boselli
1999). In the central $\sim$1.8$\times$1.5 ${\rm Mpc}^2$ region of
these two systems, 3 and 2 galaxies respectively have radio
luminosities above the adopted AGN/SF boundary (from Table 1 of
Gavazzi \& Boselli 1999).

The small number of radio galaxies did not allow us to determine the
radio luminosity function (RLF) for early-type galaxies. We compared
the number of detected radio galaxies with the result of Ledlow and
Owen (1996), who detected ~12\% of galaxies with optical magnitude
R$\leq$17.5 and radio luminosity Log $(L_{1.4~GHz}{\rm (W/Hz)})>$22.02
(scaled to our cosmological parameters). In the same radio and optical
range, we found six cluster members, of which three have clear
elliptical morphology, and two can be safely classified as AGN.  Due
to the lack of complete spectroscopy, we assumed that all the red
sequence galaxies with R$\leq$17.5 belong to A3921, finding in total
36 galaxies. Therefore, the fraction of radio emitting elliptical
galaxies in our sample is 8\%, and the fraction of AGN is 6\%.  Taking
into account the uncertainty on the total number of galaxies and the
fact that these percentages are derived from very small numbers, we
conclude that our result is not significantly different from that of
Ledlow and Owen (1996).

\section{The peculiar radio source SUMSS J224857$-$642059 
(R5).}\label{J2249-6420}

\subsection{Total intensity and spectral index images}

The most interesting radio emission observed in the central field of
A3921 is the extended source SUMSS J224857-642059 (R5) (see
Fig.~\ref{fig:grey22}). In Fig.~\ref{fig:BG2-RS5}, we have
superimposed the 1.344~GHz radio contours on the R-band optical image
of A3921 from F05. The radio component in the North-East extremity of
R5 is coincident with a very bright galaxy (${\rm R}_{AB}=16.1$)
corresponding to the second brightest cluster member (``BG2'' in F05),
whose colour and spectrum are typical of an early-type object. BG2 is
the dominant galaxy of one of the two colliding sub-clusters revealed
by the combined optical/X-ray analyses of the cluster by F05 and B05.

\begin{figure}
\begin{center}
\resizebox{9cm}{!}  
{\includegraphics{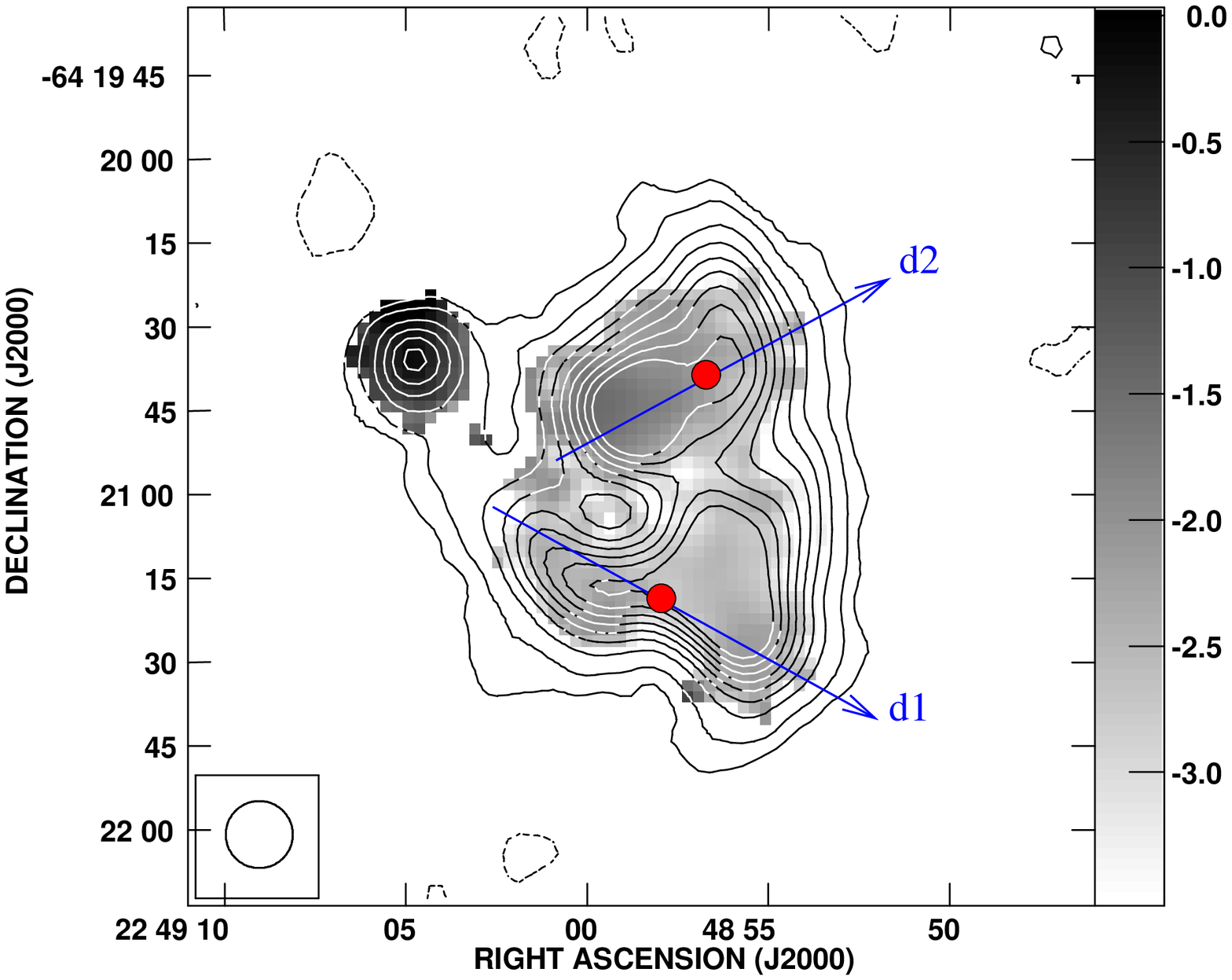}} \hfill
\parbox[b]{9cm}{
\caption{Spectral index map of the radio source SUMSS 
J224857$-$642059 (R5) made by combining the images at 1.344~GHz and
2.368~GHz. Contours are the same as in Fig.~\ref{fig:BG2-RS5}.  The
directions d1 and d2 along which the spectral index trend has been
plotted in Fig.~\ref{SpIndTrend} are also shown. Circles give the
position of the centre of the two slides of Fig.~\ref{SpIndTrend}.}
\label{SpInd}}  
\end{center}
\end{figure}

The most likely hypothesis is that R5 is a narrow-angle tail (NAT)
radio galaxy, whose core component is associated with the galaxy
BG2. A source with very similar structure has been detected by
Gregorini \& Bondi (1989) in A115. The jets in R5 are not visible at
our resolution, and the tails of diffuse emission, embedded in a
common low-brightness envelope, are oriented along a NE/SW (d1) and a
SE/NW (d2) axis (Fig.~\ref{SpInd}). The main elongation axis of the
tail (d1) is roughly aligned with the direction along which A3921-B is
tangentially traversing A3921-A (SW/NE, see
Fig.~\ref{fig:clumps}). This agrees with the accepted view that NAT
morphologies are the result of the ram-pressure exerted by the ICM on
the radio plasma, the ram-pressure being due to relative motion
between the ICM and the radio emitting galaxy (Blandford \& Rees
1974). The radial velocity offset of BG2 from the mean radial velocity
of the cluster is $-$145 $\pm$79~km/s. However, the proper velocity of
BG2 with respect to the ICM could be higher, since it has been shown
that the merging axis of A3921-B is nearly on the plane of the sky
(F05), resulting in a small velocity component along the line of
sight.

We obtained the spectral index map of the radio source R5 by combining
the images at 1.344~GHz and 2.368~GHz with the {\it AIPS} task {\it
COMB} (Fig.~\ref{SpInd}). We note a steepening of the spectrum going
from the region around BG2 to the extended emission, supporting the
classification of R5 as a tailed radio galaxy. In particular, the core
component coincident with BG2 shows a significantly flat spectral
index ($<\alpha>=-0.4$), with a steepening of the spectrum to
$\alpha=-1{\div}-1.5$ on the SW side.

\begin{figure}
\begin{center}
\resizebox{6cm}{!}  
{\includegraphics{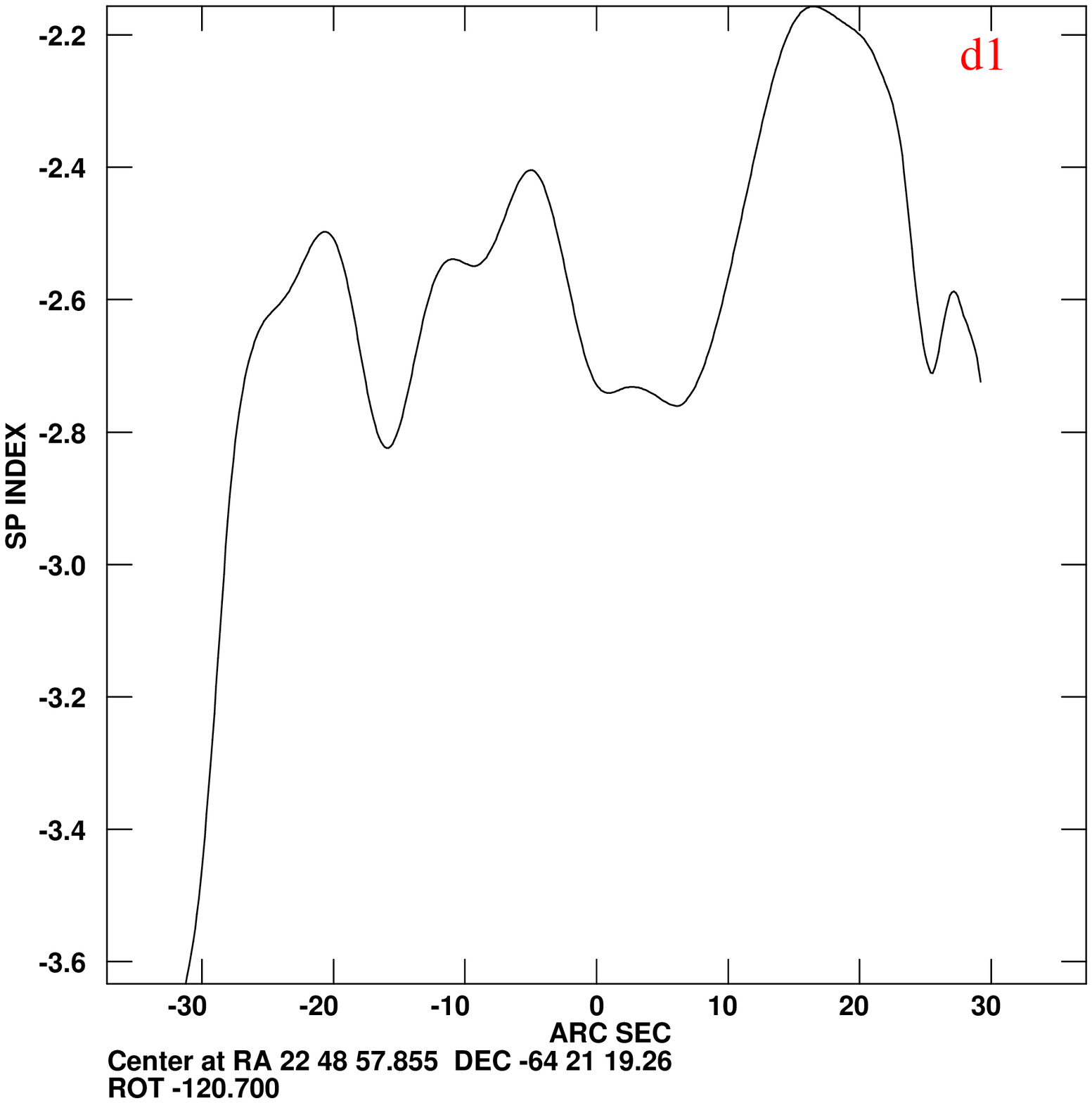}}
\resizebox{6cm}{!} 
{\includegraphics{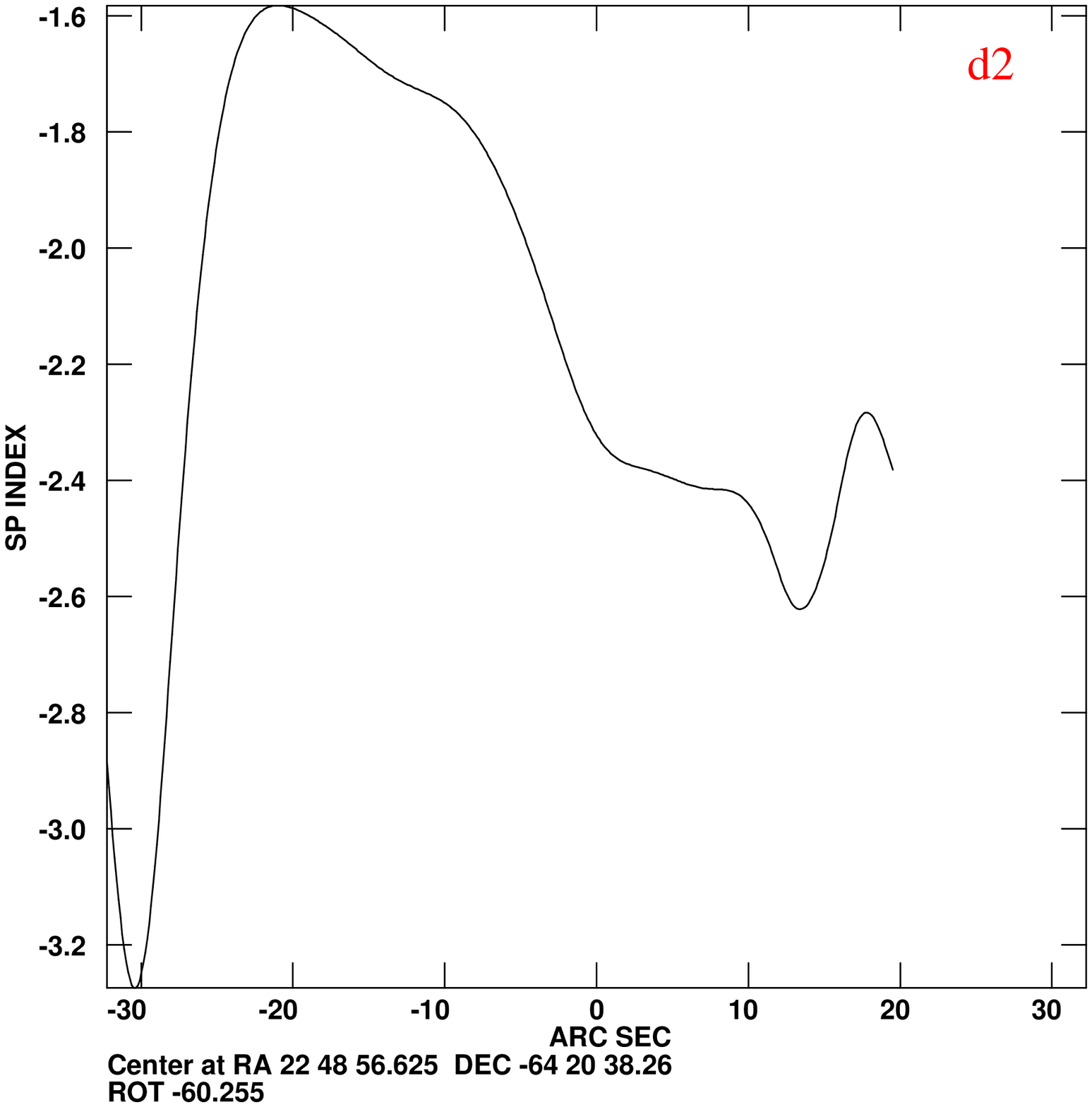}}
 \hfill  
\parbox[b]{9cm}{
\caption{Point-to-point spectral index between 1.344 and 2.368 
GHz along the main axis of SUMSS J224857$-$642059 (R5) tails (d1 and 
d2 in Fig.~\ref{SpInd}).}
\label{SpIndTrend}}  
\end{center}
\end{figure}

The trend of the spectral index of the R5 tails along the main axes
(d1 and d2 in Fig.~\ref{SpInd}) is shown in Fig.~\ref{SpIndTrend}. The
tail spectrum is very steep in both directions. We found a steepening
spectral index going away from BG2 along d2, while the spectrum shows
more jumps along d1.

The physical properties of the extended radio source R5 are 
summarised in Table~\ref{tab:J2249-6420}. Equipartition magnetic 
field was calculated with the standard assumptions of equal energy 
density in protons and electrons, a magnetic field filling factor of 
1, and a spectrum extending from 10 MHz to 100 GHz. The spectral 
index was taken $\alpha$=$-$1.5 and the source depth along the line 
of sight of 50 ${h_{75}}^{-1}$ kpc. The source age given in the 
Table refers to lifetime of electrons radiating at 1.4 GHz in a 
magnetic field taken at the equipartition value.

\begin{table}
\begin{center}   
\begin{tabular}{ll}  
\hline
\hline
Core ${\rm RA}_{\rm J2000}$ & $22^h 49^m 04^s.7$ \\
Core ${\rm Dec}_{\rm J2000}$ & $-64^{\circ} 20' 36''.6$ \\
Optical id. ${\rm RA}_{\rm J2000}$ & $22^h 49^m 04^s.8$ \\
Optical id. ${\rm Dec}_{\rm J2000}$ & $-64^{\circ} 20' 35''.6$ \\
Angular size (1.344~GHz) & $\sim 105''{\times}100''$ \\
Physical size (1.344~GHz) & $\sim 170{\times}160~{h_{75}}^{-1}~{\rm kpc}$ \\
Flux density (1.344~GHz) & 43.7 mJy \\
Power (1.344~GHz) & $8.5{\times}10^{23}~{h_{75}}^{-2}~{\rm W/Hz}$ \\
$\langle RM \rangle$ & 22.5~rad/${\rm m}^2$ \\
$\sigma_{RM}$ & 4.8~rad/${\rm m}^2$ \\
$|RM_{max}|$ & 33.5 rad/${\rm m}^2$ \\
H$_{\rm eq}$ & 3.6 $\mu$G \\
Age & 8.7${\times}10^7$ yr \\
\hline
\end{tabular}
\caption{Properties of the radio galaxy SUMSS J224857$-$642059 (R5). 
}
\label{tab:J2249-6420}
\end{center}
\end{table}

\subsection{Polarization images}

Images in the Stokes parameters U and Q were produced analogously to
the I maps of the cluster by INVERTing, CLEANing and RESTORing the
data with the {\it MIRIAD} software. In particular, for each pointing,
the ($u,v$) data at the same frequency of all the different
configurations were handled together. We restored the 2.368 GHz maps
with a beam size of both $8''{\times}8''$ (i.e. 13 cm beam size), and
$12''{\times}12''$ (i.e. 22 cm beam size). The images of each field
were then mosaiced using the task {\it LINMOS}.

Using the {\it MIRIAD} task {\it IMPOL}, we derived from the I, Q and
U maps the images of: a) the polarised intensity
$P~=~(Q^2~+U^2)^{1/2}$, b) the degree of polarization $m~=~P/I$, c)
the position angle of polarization (defined to be positive North
through East) $\psi~=~0.5~{\rm tan}^{-1}(U/Q)$, and d) the associated
errors. These maps were all blanked when the error in the position
angle image was greater than $10^{\circ}$ or when the $S/N$ ratio in
the polarised intensity map was less than three.

The source R5 is polarized in the Northern region both at 1.344 and
2.368 GHz. In the radio images presented in Fig.~\ref{fig:BG2-RS5} the
vectors indicate the orientation of the projected {\it {\bf E}}-field
and are proportional in length to the fractional polarization.  The
mean fractional polarization is $\sim$11\% and $\sim$40\% respectively
at 1.344 and 2.368 GHz, reaching values of $\sim$34\% and $\sim$60\%
at the Northern boundary of the source, whereas for the Southern tail
we can set upper limits of $\sim$6\% and $\sim$20\% at 1.344 and 2.368
GHz respectively. A possible interpretation of the different
polarization percentage in the two tails is that the source is
strongly affected by projection effects.

As the polarized emission from a radio source passes through a
magnetized, ionized plasma, its plane of polarization will be rotated
by the Faraday effect:

\begin{equation}
\psi_{\lambda}~=~\psi_0~+~RM~\lambda^2,
\label{eqn:RM1}
\end{equation}

\noindent where $\psi_{\lambda}$ and $\psi_0$ are the observed and 
intrinsic polarization angles respectively and $\lambda$ is the
wavelength at which observations are performed. $RM$ is the Rotation
Measure, which depends on the product of the electron density ($n_e$)
and magnetic field along the line of sight ($H_{\|}$). Due to this
effect, the plane of polarized radiation from cluster and background
radio galaxies may be rotated if magnetic fields are present in the
ICM.

We derived the rotation measure ($RM$) and the zero-wavelength 
position angle ($\psi_0$) maps by fitting Eq.~\ref{eqn:RM1} to the 
position angle images of the radio galaxy R5 at 1.344 GHz and 2.368 
GHz with a resolution of 12.0$''$. For this, we used the {\it 
MIRIAD} task {\it IMRM}. The measured RM values range from 11.2 to 
33.5 rad/${\rm m}^2$. They are consistent with the Galactic 
contribution, which in the region of A3921 is expected to be about 
30 rad/${\rm m}^2$ (Simard-Normandin et al. 1981). This suggests 
that the cluster magnetic field in this region may be low or mostly 
perpendicular to the line of sight. The orientation of the magnetic 
field of R5 is roughly longitudinal as commonly found in tailed 
radio sources (Dallacasa et al. 1989; Feretti et al. 1998).

\section{Discussion and conclusions}\label{disc}

\subsection{Star-forming galaxies and AGNs: possible connections with 
the ongoing merger}

In this paper we presented the 22 and 13 cm ATCA observations of the
galaxy cluster A3921. We concentrated our analysis on the central
field of the cluster (0.075 deg$^2$), which hosts two merging
sub-clusters (A3921-A and A3921-B) that have not yet undergone the
first core-core encounter (B05; F05; Kapferer et al. 2006). Our main
aim was to closely investigate the possible connection between the
on-going merging event and the SF properties of cluster members
suggested by F05. We detected 17 radio sources, among which seven are
cluster galaxies (confirmed or candidate). Based also on their optical
and X-ray properties, the latter were classified as: a) two AGNs (R5
and R9), b) four late-type, gas rich galaxies (R2, R10, R11 and R14),
and c) one (R3) strong star-burst.

Seven radio emitting galaxies were therefore detected in the 
central field of A3921 at our 5$\sigma$ sensitivity level, of
which four are late-type objects. In an equivalent cluster region 
and at the same flux limit (scaled to the proper $z$), 12 galaxies 
were detected at 1.4 GHz both in A1367 and in Coma, of which nine 
and seven, respectively, have late-type morphologies (from Table 1 
by Gavazzi \& Boselli 1999). Assuming Poissonian noise, we thus 
found a slight deficiency (${\sim}1\sigma$ level) of radio galaxies 
in A3921 with respect to the lower-$z$ merging clusters Coma and 
A1367.

Among the 11 emission-line galaxies identified by F05, only two (the
previously mentioned R2 and R3) were detected at our 3$\sigma$
sensitivity level (Log $(L_{22 cm}{\rm (W/Hz)})\sim$21.4. The
remaining nine objects are located in the collision region of the
merging cluster. The lack of radio emission gave us an upper limit for
their SFR unbiased by dust extinction (SFR$\lesssim3 {{\rm
M}_{\odot}{\rm yr}^{-1}}$). Based on their observational properties
(optical and radio luminosity, colour, presence and strength of
emission lines in their spectra, SFR) and on their position in the
cluster, we thus suggested that most of the emission-line galaxies
detected in the collision region of A3921 are not infalling, gas-rich
objects. Their SF episode would instead be related to the on-going
sub-cluster collision. The merging event could have favoured both
ICM-galaxy and tidal (galaxy-galaxy and cluster-galaxy) interactions,
which have been shown to be able to trigger SF (e.g. Evrard 1991; Moss
\& Whittle 2000; Poggianti et al.  2004; Bekki 1999; Kapferer et
al. 2005; Boselli \& Gavazzi 2006).  The cluster merger is probably
also responsible for the strong star-burst in R3 (BG3 in F05 and B05):
the compression of the ICM on the ISM of this object is probably very
strong, due both to the high ICM density (it is located at the X-ray
centre of A3921-B, B05) and to the high proper velocity of this object
(F05).

Note that, excluding R3, we did not detect any gas-rich, 
star-forming galaxy in the collision region 
(Fig.~\ref{fig:FOC_radio_ottico} and Table~\ref{tab:SFR}). As 
summarised at the beginning of this Sect. (point b)), four gas-rich, 
star-forming galaxies have been detected through our radio 
observations. All of them are located in the external parts of the 
merging sub-clusters, and they probably are infalling, field 
spirals, whose gas reservoir has not yet been emptied by cluster 
environmental effects (Dressler \& Gunn 1983).

The absence of radio emission in all the k+a galaxies detected in
A3921 confirms that they are not dusty star-bursting objects. This
result agree with the conclusion of F05 and of previous analyses
(e.g. Duc et al. 2002) that k+a objects are observed during a
post-star-forming phase. It also confirms that, in the A3921 central
field, star-forming objects are concentrated in the collision region.

Our counts of radio emitting elliptical galaxies are not significantly
different from the general cluster RLF derived by Ledlow and Owen
(1996). Similar numbers of early-type objects with radio luminosity
typical of AGNs were detected in two other low-$z$ clusters, i.e. Coma
and A1367 (Gavazzi \& Boselli 1999).  Analogous results were also
found by Venturi et al. (2001) in the pre-merging cluster A3528,
suggesting that, in their early stages, cluster mergers do not
influence the emission from radio-loud AGNs.

\subsection{The peculiar radio galaxy SUMSS J224857-642059 (R5)}

In order to understand the nature of the radio galaxy SUMSS 
J224857-642059 (R5) and the origin of its peculiar morphology, we 
have to take into account the following observational evidence:

\begin{itemize}

\item the position of its possible core component, coincident at $\sim$ 1 
arcsec uncertainty with the position of the optical galaxy BG2 (see
Tables \ref{tab:ID} and \ref{tab:J2249-6420});

\item its location at the centre of the less massive of two merging 
sub-clusters (A3921-B);

\item its distorted tailed radio emission, without collimated jets, 
and roughly aligned with the merging axis (SW/NE). A NW elongation 
is, however, also present, and this shows a high polarized 
fraction and a nearly transverse intrinsic magnetic field;

\item its steep total spectral index ($<\alpha 
^{2.368}_{1.344}>$=$-$2.4), typical of tailed radio sources in 
clusters of galaxies;

\item the significantly flatter spectrum of the component associated 
with BG2 with respect to the extended emission ($<\alpha>=-0.4$ and
$<\alpha>=-2.6$).

\end{itemize}

\noindent We therefore consider that R5 is a NAT radio 
source associated with BG2, with a core component in its 
Northernmost
side, and an extended ($\sim170{\times}160$~\h~kpc) tail. This latter
is distorted towards NW, possibly due to a lower local density of the
intra-cluster gas revealed by the XMM observations of B05. The tail
morphology is very likely explained by ram-pressure effects due to the
relative motion between the ICM and BG2, in turn related to the
merging event of the host sub-cluster (A3921-B). The total radio power
of R5 at 1.344~GHz is Log $(P{\rm (W/Hz)})$=24.0. Its total flux
density is dominated by the extended emission located outside the
optical galaxy, with a flux density ratio of $\sim$18 between the
extended and the core component ($\sim41.4$ and $\sim2.3$ mJy
respectively).

R5 does not show visible jets and the emission in the tail is very 
steep ($\alpha{\simeq}-2{\div}-3.5$), suggesting that electrons are 
suffering strong radiation losses. The diffuse component of R5 would 
therefore be a `dying' tail, which has been partly detached 
from an earlier period of activity of the BG2 radio galaxy and whose 
final evolution is dominated by synchrotron losses. The extended 
emission is now presumably supported and distorted by the ICM 
pressure. The strong polarization at the extreme of the NW component 
of the tail suggests that the radio jet has encountered a denser (or 
higher pressure) region leading to magnetic field compression.

\begin{acknowledgements}  
CF is very grateful to Lister Staveley-Smith for his valuable help in
ATCA data reduction. We warmly thank Vincent McIntyre for having
provided the new version of the $MIRIAD$ script ``quickfit''. We thank
Christophe Benoist, Wolfgang Kapferer, Rocco Piffaretti, Alberto
Cappi, Pierre-Alain Duc, Elaine Sadler and Jean-Luc Sauvageot for very
fruitful discussions. We are indebted to the referee, Giuseppe
Gavazzi, for useful suggestions which helped to improve the paper.  CF
acknowledges financial support of Sydney University, Australia
Telescope National Facility, Marie Curie individual fellowship
MEIF-CT-2003-900773, Austrian Science Foundation (FWF) through grant
number P15868, and Tiroler Wissenschaftsfonds (TWF). The Australia
Telescope Compact Array is part of the Australia Telescope which is
funded by the Commonwealth of Australia for operation as a National
Facility managed by CSIRO.
\end{acknowledgements}

{}
\end{document}